\documentclass[a4paper,fleqn,usenatbib]{mnras}

\usepackage{mathptmx}

\usepackage[T1]{fontenc}
\usepackage{ae,aecompl}

\usepackage{graphicx}
\usepackage{amsmath}	
\usepackage{amssymb}	
\usepackage{mathrsfs}
\usepackage{url}
\usepackage{bm}
\usepackage{booktabs}
\usepackage{multicol}
\usepackage{comment}

\newcommand{\msun}{\ensuremath{\,\textrm{M}_{\odot}}}

\newcommand{\mzams}{\ensuremath{\,M_{\textrm{ZAMS}}}}
\newcommand{\bse}{{\sc BSE}}
\newcommand{\mobse}{{\sc MOBSE}}
\newcommand{\sevn}{{\sc SEVN}}

\title[Merging black hole binaries]{Merging black hole binaries: the effects of progenitor's metallicity, mass-loss rate and Eddington factor}

\author[N. Giacobbo et al.]{
Nicola Giacobbo,$^{1,2}$\thanks{E-mail: nicola.giacobbo@oapd.inaf.it}
Michela Mapelli,$^{2,3,4}$
Mario Spera,$^{2,3,4}$
\\
$^{1}$Dipartimento di Fisica e Astronomia ``G. Galilei'',
    Universit\`a di Padova, vicolo dell'Osservatorio 3, I-35122 \\
$^{2}$INAF, Osservatorio Astronomico di Padova, vicolo dell'Osservatorio 5, I--35122 Padova, Italy.\\
$^{3}$Institute for Astrophysics and Particle Physics, University of Innsbruck, Technikerstrasse 25/8, A--6020, Innsbruck, Austria\\ 
$^{4}$INFN, Milano Bicocca, Piazza della Scienza 3, I--20126, Milano, Italy\\
}

\date{Accepted XXX. Received YYY; in original form ZZZ}

\pubyear{2017}

\begin{document}
\label{firstpage}
\pagerange{\pageref{firstpage}--\pageref{lastpage}}
\maketitle

\begin{abstract}
The first four gravitational wave events detected by LIGO were all interpreted as merging black hole binaries (BHBs), opening a new perspective on the study of such systems. Here we use our new population-synthesis code \mobse{}, an upgraded version of \bse{} \citep{Hurley2002}, to investigate the demography of merging BHBs. \mobse{} includes metallicity-dependent prescriptions for mass loss of massive hot stars. It also accounts for the impact of the electron-scattering Eddington factor on mass loss. We perform $>10^8$ simulations of isolated massive binaries, with 12 different metallicities, to study the impact of mass loss, core-collapse supernovae and common envelope on merging BHBs. Accounting for the dependence of stellar winds on the Eddington factor leads to the formation of black holes (BHs) with mass up to 65 M$_\odot$ at metallicity $Z\sim{}0.0002$. However, most BHs in merging BHBs have masses $\lesssim{}40$ M$_\odot$. We find merging BHBs with mass ratios in the $0.1 - 1.0$ range, even if mass ratios $>0.6$ are more likely. We predict that systems like GW150914, GW170814 and GW170104 can form only from progenitors with metallicity $Z\leq{}0.006$, $Z\leq{}0.008$ and $Z\leq{}0.012$, respectively. Most merging BHBs have gone through a common envelope phase, but up to $\sim{}17$ per cent merging BHBs at low metallicity did not undergo any common envelope phase. We find a much higher number of mergers from metal-poor progenitors than from metal-rich ones: the number of BHB mergers per unit mass is $\sim{}10^{-4}$ M$_\odot^{-1}$ at low metallicity ($Z = 0.0002 - 0.002$) and drops to $\sim{}10^{-7}$ M$_\odot^{-1}$ at high metallicity ($Z \sim{} 0.02$).
\end{abstract}




\section{Introduction}
The first four direct detections of gravitational waves (GWs, \citealt{Abbott2016a,Abbott2016b,Abbott2016c,Abbott2017,LIGO2017}) revolutionised our knowledge of black hole binaries (BHBs). Thanks to them, we now know that coalescing BHBs exist and can host massive black holes (BHs), with mass $\gtrsim{}30$ M$_\odot$ (as in the case of GW150914, GW170104 and GW170814).

The formation and evolution of BHBs have been investigated for a long time \citep[e.g.][]{Tutukov1973,Thorne1987,Schutz1989,Kulkarni1993,Sigurdsson1993,Portegies2000,Colpi2003,Belczynski2004}. Already before the first LIGO detection, several theoretical studies predicted the existence of stellar BHs with mass $\gtrsim{}30$ M$_\odot$ \citep[e.g.][]{Heger2003,Mapelli2009,Mapelli2010,Belczynski2010,Fryer2012,Mapelli2013,Spera2015}. The basic idea is that if the pre-supernova mass of a star is sufficiently large \citep{Fryer1999,Fryer2001} and/or its pre-supernova compactness sufficiently high \citep{Oconnor2011}, the star can collapse to a BH directly, producing a more massive remnant than in case of a supernova (SN) explosion.

Unfortunately, state-of-the-art theoretical models of BHBs still suffer from major uncertainties. The main issues are the treatment of core-collapse supernovae, stellar winds, and common envelope (CE, e.g. \citealt{Dominik2012}). The impact of stellar dynamics on the formation of BHBs is also matter of debate (e.g. \citealt{Ziosi2014,Rodriguez2015,Rodriguez2016,Mapelli2016,Askar2016,Antonini2017,Banerjee2017}).

 The physics of core-collapse SNe is remarkably complex and barely understood \citep[e.g.][]{Fryer1999,Heger2002,Heger2003,Fryer2006,Oconnor2011,Fryer2012,Janka2012,Burrows2013, Pejcha2015,Woosley2017}. The key point is to understand the connection between the final stages of a massive star's life and the outcome of a SN \citep[for a review, see][]{Limongi2017}. In particular, it is crucial to assess what are the conditions for a star to directly collapse into a BH, without an explosion \citep[e.g.][]{Bethe1990, Fryer1999,Fryer2001,Janka2007, Oconnor2011, Janka2012, Burrows2013, Ertl2016}.

Mass loss by stellar winds is also crucial, because it governs the final mass $M_{\rm{fin}}$ of a star, i.e. the mass of a star just before the SN \citep[see e.g.][]{Mapelli2009,Belczynski2010,Fryer2012,Mapelli2013}. 
In the last decade, models of line-driven stellar winds were profoundly revised.  Current models suggest a strong dependence of mass loss on metallicity ($\dot{M}\propto{}Z^\alpha{}$, with $\alpha{}\sim{}0.85$, \citealt{Vink2001,Vink2011,Muijres2012}) not only during the main sequence (MS) but also after, including the Wolf-Rayet (WR) stage \citep[see e.g.][]{VinkdeKoter2005,Meynet2005,Graefener2008,Vink2011,Tang2014,Chen2015}.

 In addition, there has been a debate regarding the importance of stationary versus eruptive mass loss for massive star evolution \citep{Vink2012}. Clumping of stellar winds results in a reduction of the mass-loss rate. Recent theoretical models (e.g. \citealt{Vink2001}) predict lower mass-loss rates by a factor of 2 -- 3 with respect to unclumped empirical mass-loss rates, consistent with a moderate wind clumping. 

Finally, some of the most recent wind models \citep[e.g.][]{Graefener2008,Graefener2011,Vink2011,Vink2016} and observations \citep[e.g.][]{Bestenlehner2014} suggest that mass loss also depends on the electron-scattering Eddington factor $\Gamma_e$ of the star. In particular, if a star  approaches the Eddington limit (i.e. $\Gamma_e\lesssim{}1$), stellar winds become almost insensitive to metallicity. Most stellar evolution models do not include this dependence, with very few exceptions (e.g. {\sc PARSEC}, \citealt{Chen2015}). Population-synthesis codes, which are used to study the demography of BHs and BHBs, should also account for these updated models of stellar winds and massive star evolution.

Currently, only few population-synthesis codes adopt up-to-date metallicity-dependent prescriptions for stellar winds  \citep[e.g.][]{Belczynski2010,Toonen2012,Mapelli2013,Spera2015,Spera2017,Banerjee2017}. The dependence of stellar winds on the Eddington factor is implemented only in the \sevn{} code (where \sevn{} is the acronym for Stellar EVolution for N-body, \citealt{Spera2015,Spera2017}). In its published version, \sevn{} only evolves single stars and is currently undergoing a major upgrade to include the main binary evolution processes (Spera et al., in preparation).

Here we present our upgraded version of \bse{} (acronym for Binary Stellar Evolution), one of the most popular population-synthesis codes  \citep{Hurley2000,Hurley2002}. In the following, we refer to our new version of \bse~as \mobse~(which stands for `Massive Objects in Binary Stellar Evolution'). 
With respect to \bse{}, \mobse{} includes up-to-date equations for metal-dependent stellar winds (based on \citealt{Belczynski2010} and on \citealt{Chen2015}) and new prescriptions for core-collapse SNe (based on \citealt{Fryer2012}). Moreover, \mobse{} includes the dependence of stellar winds on the Eddington factor, adopting the prescriptions by \cite{Chen2015}.

We use \mobse{} to investigate the formation of BHBs from isolated binaries, comparing different models of stellar winds and SNe. We also examine the role played by the CE phase in the formation of BHBs.

\section{Methods}
\label{sec:2}
In this Section, we describe the main features of {\sc MOBSE} with respect to {\sc BSE}. 

\subsection{Stellar winds and mass loss}
\label{sec:2.1}

\mobse{} adopts the following prescriptions for stellar winds. For O and B stars with effective temperature $12500\,{}{\rm K}\leq{}T\leq{}25000\,{}{\rm K}$ we adopt equation 25 of \citet{Vink2001}:
\begin{eqnarray}\label{eq:vink25}
 \log{ \dot{M}} =& -6.688 + 2.210\,{}\log{\left(\frac{L}{10^5~\text L_{\odot}}\right)} \nonumber \\
 & -1.339\,{}\log{\left(\frac{M}{30~\text M_{\odot}}\right)} - 1.601\,{}\log{\left(\frac{V}{2.0}\right)} \nonumber \\
 & +\alpha{}\,{}\log{\left(\frac{Z}{\text Z_{\odot}}\right)} + 1.07\,{}\log{\left(\frac{T}{20000~ {\rm K}}\right)}, 
\end{eqnarray}
where $L$ is the stellar luminosity, $M$ is the stellar mass, $\alpha{}$ expresses the dependence of mass loss on metallicity and $V=v_{inf}/v_{esc} = 1.3$ is the ratio of the wind velocity at infinity ($v_{\text{inf}}$) to escape velocity ($v_{\text{esc}}$).

For O and B stars stars with $25000~ {\rm K} < T \leq 50000~ {\rm K}$ we use equation 24 of \citet{Vink2001}:
\begin{eqnarray}\label{eq:vink24}
 \log{ \dot{M}} =& -6.697 + 2.194\,{}\log{\left(\frac{L}{10^5L_{\odot}}\right)} \nonumber \\
 & -1.313\,{}\log{\left(\frac{M}{30~\text M_{\odot}}\right)} - 1.226\,{}\log{\left(\frac{V}{2.0}\right)} \nonumber \\
 & +\alpha{}\,{}\log{\left(\frac{Z}{\text Z_{\odot}}\right)} + 0.933\,{}\log{\left(\frac{T}{40000~ {\rm K}}\right)} \nonumber \\
 & 10.92\,{}\left[\log{\left(\frac{T}{40000~ {\rm K}}\right)}\right]^2, 
\end{eqnarray}
where $V=v_{inf}/v_{esc} = 2.6$. The above dichotomy is due to the bi-stability jump, i.e. a sudden jump in the mass-loss rate related to the fact that the iron ions driving the wind recombine at $T\sim{}25000$ K, and again below 12500 K \citep[for more details see][]{Vink1999,Petrov2016}.

We express the mass loss of luminous blue variable (LBV) stars as
\begin{equation}\label{eq:LBV}
  \dot{M} = 10^{-4}\,{}f_{\rm LBV}\,{}\left(\frac{Z}{\text Z_{\odot}}\right)^{\alpha{}}\,{} {\rm M}_{\odot}\,{}{\rm yr}^{-1},
\end{equation}
 where $f_{\rm LBV}$ is a parameter (we choose $f_{\rm LBV}=1.5$, in agreement with \citealt{Belczynski2010}).

Finally, for Wolf-Rayet (WR) stars we use equation 9 of \citet{Belczynski2010}:
\begin{equation}\label{eq:WR}
	\dot{M}_{\text{WR}} = 10^{-13}L^{1.5}\,{}\left(\frac{Z}{\text Z_{\odot}}\right)^{\alpha{}} ~{\text{M}}_{\odot}~{\text{yr}}^{-1}~.
\end{equation}

For the other stars, {\sc mobse} adopts the same mass loss formalism as the original version of {\sc bse}. 

\subsubsection{\sc mobse1}
Equations~\ref{eq:vink25}, \ref{eq:vink24}, \ref{eq:LBV} and \ref{eq:WR}, contain the parameter $\alpha{}$ which expresses the dependence of mass loss on metallicity. In our fiducial version of {\sc mobse} (hereafter, {\sc mobse1}), we define $\alpha{}$ in the following way:

\begin{equation}
\label{eq:scaling}
\alpha = \begin{cases} 0.85 & \rm{if}~~~ \Gamma_{\rm e} < 2/3 \cr 2.45 - 2.4 \,{} \Gamma_{\rm e} & \rm{if}~~~ 2/3 \leq \Gamma_e \leq 1~, \cr
\end{cases}
\end{equation}
where $\Gamma_e$ is expressed as (see equation~8 of \citealt{Graefener2011}):
\begin{equation}
\log{\Gamma_e}=-4.813+\log{(1+X_{\rm H})}+\log{(L/L_\odot)}-\log{(M/M_\odot)},
\end{equation}
where $X_{\rm H}$ is the Hydrogen fraction.

According to this definition, the dependence of mass loss on metallicity almost vanishes when the star is radiation pressure dominated ($\Gamma_e\sim{}1$). This expression for $\alpha{}$ was derived by \citet{Tang2014} and \citet{Chen2015}, based on the results of \cite{Graefener2008}. In fact, \cite{Graefener2008} and \cite{Vink2011} show that the mass-loss rate  is strongly enhanced when the star approaches the electron-scattering Eddington limit $\Gamma_e$. This means that increasing $\Gamma_e$ the metallicity dependence becomes weaker.

\subsubsection{\sc mobse2}
Other population-synthesis codes (e.g. {\sc STARTRACK}, \citealt{Belczynski2010}) do not take into account the effect of $\Gamma_e$ on mass loss. To quantify the importance of $\Gamma_e$ and to compare our results with previous work, we also introduce a second version of our {\sc mobse} code (which we will refer to as {\sc mobse2}), where we do not include the effect of $\Gamma_e$. 

In {\sc mobse2}, the parameter $\alpha{}$ is defined as

\begin{equation}
\label{eq:scaling2}
\alpha = \begin{cases} 0.85 & \rm{in\,{}equations~\ref{eq:vink25}\,{}and~\ref{eq:vink24}} \cr 0 & \rm{in\,{}equation~\ref{eq:LBV}} \cr 0.86 & \rm{in\,{}equation~\ref{eq:WR}}.\cr
\end{cases}
\end{equation}

The values of $\alpha{}$ defined in equation~\ref{eq:scaling2} are the same as adopted by \cite{Belczynski2010}. This implies that in {\sc MOBSE2} mass loss of O and B-type stars scales as $\dot{M}\propto{}Z^{0.85}$ \citep{Vink2001}. WR stars also show a similar dependence ($\alpha{}\sim{}0.86$, \citealt{VinkdeKoter2005}). Finally, the mass loss of LBVs  does not depend on metallicity, as in equation~8 of \cite{Belczynski2010}. This means that mass loss of LBVs is constant, while mass loss of MS and WR stars does not depend on $\Gamma_e$.

\subsection{Supernovae}
\label{sec:2.2}
The physics of core-collapse SNe  is  uncertain and several different models exist \citep[see][for a review]{Smartt2009}. For this reason, we have implemented two different prescriptions for core-collapse SNe (described in details by \citealt{Fryer2012}): i) the \textit{rapid} SN model and ii) the \textit{delayed} SN model.

The main difference between them is that they assume a different time-scale at which the explosion occurs: in the rapid (delayed) model the explosion takes place  $t < 250$ ms  ($t \gtrsim 0.5$ s) after core bounce.

Both the prescriptions depend only on the final mass of the Carbon-Oxygen (CO) core ($M_{\rm CO}$) and on the final mass of the star ($M_{\rm fin}$), which determines the amount of fallback. 

Other studies suggest a more complex relation between the properties of the star at the onset of collapse and the compact remnant mass (e.g. \citealt{Oconnor2011,Janka2012,Ugliano2012,Pejcha2015,Ertl2016}) and provide alternative formalisms to predict the remnant mass (e.g. the compactness criterion by \citealt{Oconnor2011} or the two-parameter criterion by \citealt{Ertl2016}). However, we cannot  adopt these alternative formalisms in {\sc BSE}, because they rely on the inner structure of the star at the onset of collapse, which is not calculated in {\sc BSE}. Figure~21 of the recent review by \cite{Limongi2017} shows that there is a strong correlation between the compactness parameter and the CO mass of the progenitor star, suggesting that our formalism based on the CO mass should give results similar to the formalism based on the compactness parameter in most cases (see also Figures 21 and 22 of \citealt{Spera2015}).

Furthermore, the rapid and delayed SN models do not distinguish between neutron stars (NSs) and BHs, because they are general prescriptions for the formation of compact remnants. According to the Tolman-Oppenheimer-Volkoff limit \citep{Oppenheimer1939}, we assume that the minimum mass for a BH is $3.0\msun$ and all compact SN remnants with mass $<3.0\msun$ are NSs. 

The details of the implementation of core-collapse SNe in {\sc mobse} can be found in Appendix \ref{sec:appA}.

In {\sc MOBSE}, we also added a formalism for pair-instability SNe (PISNe, \citealt{Ober1983,Bond1984,Heger2003,Woosley2007})  and pulsational pair-instability SNe (PPISNe, \citealt{Barkat1967,Woosley2007,Chen2014,Yoshida2016}), which are not included in {\sc BSE} and in most population-synthesis codes. Our description of PISNe is based on the results by \cite{Heger2003}: if the final Helium core mass ($M_{\rm He,f}$) of a star is $64\,{}{\rm M}_\odot\lesssim{}M_{\rm He,f}\lesssim{}135\,{}{\rm M}_\odot$, we assume that the star leaves no remnant, because the ignition of Oxygen and Silicon releases enough energy to disrupt the entire star. If $M_{\rm He,f}>135$ M$_\odot$, the star is expected to avoid the PISN and to directly collapse into a BH.

Our description of PPISNe is based on the formalism presented in \cite{Spera2016} and \cite{Spera2017} (see also \citealt{Belczynski2016pair,Woosley2017}).  If the final mass of the Helium core is $32\,{}{\rm M}_\odot\lesssim{}M_{\rm He,f}\lesssim{}64\,{}{\rm M}_\odot$, the star undergoes a PPISN and leaves a compact remnant whose mass is described by the fitting formulas in Appendix~\ref{sec:appB}. 

We also updated the natal kick for BHs  ($V_{\rm kick}$) as follows \citep{Fryer2012}:
\begin{equation}
	V_{\rm kick} = (1 - f_{\rm fb})\,{}W_{\rm kick},
\end{equation}
where $f_{\rm fb}$ is the fallback factor (the explicit expression can be found in Appendix \ref{sec:appA}). $W_{\rm kick}$ is randomly drawn from a Maxwellian distribution with a one dimensional root-mean square $\sigma{}=265$ km s$^{-1}$. This distribution was derived by   \cite{Hobbs2005}, based on the proper motions of 233 isolated Galactic pulsars.

\subsection{Additional changes in \mobse}
\label{sec:2.3}

\begin{figure}
	\centering		
	\includegraphics[scale=0.36]{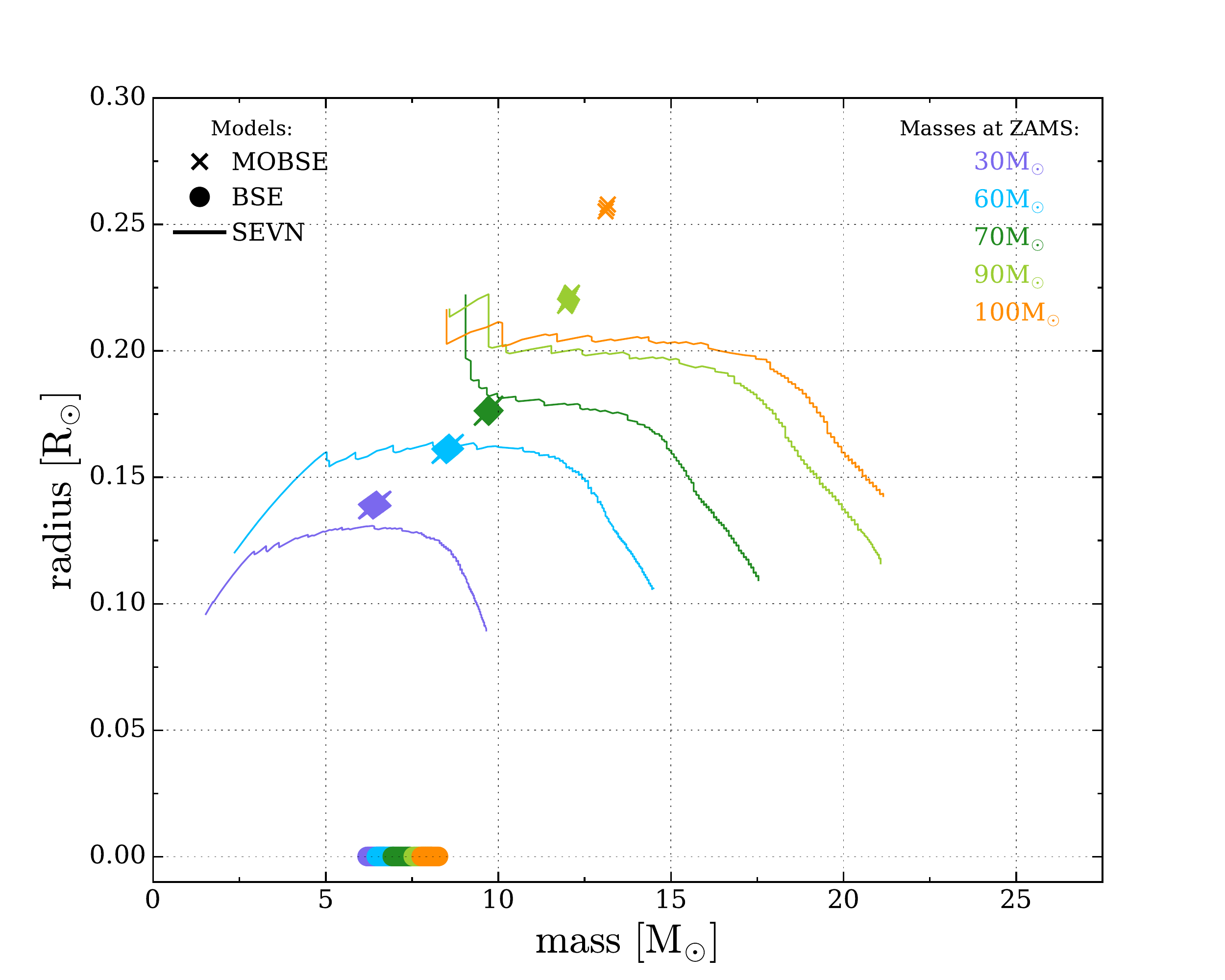}
	\caption{Radius versus mass of the CO core for five stars with zero-age main sequence (ZAMS) mass $M_{\rm{ZAMS}} =30,60,70,90,100 \msun$, at solar metallicity ($Z = 0.02$). Solid line: \sevn{}; crosses: \mobse{}; circles: \bse{}.
	\label{fig:core}}
\end{figure}	

We also added to \mobse{} the fitting formulas described in  \cite{Hall2014}, to compute the core radius of evolved stars. The core radius is crucial to determine the final fate of a CE system. Figure \ref{fig:core} shows the differences between the CO core radii computed with \bse~(triangles), \mobse~(circles) and \sevn{} (lines) as a function of the core mass. The CO core radii derived with \mobse{} are of the same order of magnitude as those obtained with \sevn{}, while \bse{} predicts unphysically small CO core radii. Our treatment of WR and LBV stars does not account for envelope inflation, which might increase the effective photospheric radii by a factor of $\sim{}10$ \citep{Graefener2012}. We will include this effect in forthcoming studies.

Another critical issue of CE  is the treatment of Hertzsprung Gap (HG) donors \citep[e.g][]{DeMink2015}. During the HG phase, stars do not present a steep density gradient between the core and the envelope and for this reason their response to the CE phase should be similar to that of MS stars \citep[see][]{Ivanova2004}. In \bse{}, when a MS star enters a CE phase as donor it merges with the accretor, while HG donors are allowed to survive the CE phase. On the contrary, in \mobse{} we imposed that even HG donors merge with their companions if they enter a CE phase \citep{Dominik2012}.

Finally, we also extended the mass range of {\sc BSE} to include stars up to $150~\msun$. Because of the fitting formulas by \cite{Hurley2000} might be inaccurate for very massive stars ( $> 100~\msun$) we imposed that the values of the stellar radii of single star are consistent with {\sc PARSEC} stellar evolution tracks \citep{Chen2015}, as discussed in \cite{Mapelli2016}. We do not consider stars with zero-age main sequence (ZAMS) mass $\mzams>150$ M$_\odot$, because the mismatch between {\sc BSE} fitting formulas and {\sc PARSEC} tracks increases significantly above this mass (see e.g. Fig.~\ref{fig:stellarwinds}).

\subsection{Simulations and initial distributions}
\label{sec:2.4}


In this section we detail the initial conditions of our population synthesis simulations. The mass of the primary star ($m_{\mathrm{1}}$) is randomly extracted from a Kroupa initial mass function \citep{Kroupa2001},
\begin{equation}
	\mathfrak{F}(m_1) ~\propto~ m_1^{-2.3} \qquad \mathrm{with}~~ m_1 \in [5-150]\msun ~.
\end{equation}
We sampled the mass of the secondary $m_{\mathrm{2}}$ according to the distribution proposed by \citet{Sana2012} 
\begin{equation}
	\mathfrak{F}(q)~ \propto ~q^{-0.1} \qquad ~~~\mathrm{with}~~~q = \frac{m_2}{m_1}~ \in [0.1-1]~m_{\mathrm{1}}~,
\end{equation}
We adopt the distributions proposed by \cite{Sana2012} also for the orbital period $P$ and the eccentricity $e$:
\begin{equation}
	\mathfrak{F}(\mathscr{P}) ~\propto~ (\mathscr{P})^{-0.55} ~~\mathrm{with}~ \mathscr{P} = \mathrm{log_{10}}(P/\mathrm{day}) \in [0.15-5.5]
\end{equation} 
and 
\begin{equation}
	\mathfrak{F}(e) ~\propto ~e^{-0.42} \qquad ~~\mathrm{with}~~~ 0\leq e < 1~.
\end{equation} 

For the CE phase we adopted the same formalism used by \citet{Hurley2002} and described in detail in \citet{Ivanova2013}. This formalism depends on two free parameters, $\alpha$ and $\lambda$, where, $\alpha$ is the fraction of  orbital energy which can be used to unbind the envelope and  $\lambda$ describes the geometry of the envelope. In this work, we consider three different combinations of these parameters: $\alpha=1.0$, $\lambda=0.1$ (which is well motivated for massive stars, see e.g. \citealt{Xi2010,Loveridge2011}), $\alpha=3.0$, $\lambda=0.5$ and $\alpha=0.2$, $\lambda=0.1$.

We ran eight sets of simulations (see Table~\ref{tab:simulations}) in order to test different combinations of stellar wind models, SN explosion mechanisms and values of $\alpha$ and $\lambda$.

For each set of simulations we performed 12 sub-sets with different metallicities $Z=0.0002$, $0.0004$, $0.0008$, $0.0012$, $0.0016$, $0.002$, $0.004$, $0.006$, $0.008$, $0.012$, $0.016$ and $0.02$. The polynomial fitting formulas implemented in {\sc BSE} \citep{Hurley2000} and the prescriptions for mass loss adopted in {\sc MOBSE} have been shown to hold in this metallicity range (e.g. \citealt{Kudritzki2002,Bresolin2004}). In each sub-set, we simulate $10^7$ binary systems. Thus, each of the eight sets of simulations is composed of $1.2\times10^8$ massive binaries.

\begin{table}
	\begin{center}
		\caption{Initial conditions.\label{tab:simulations}}
		\begin{tabular}{ccccc}
			\toprule
			ID & Winds &  SN &  $\bm{\alpha }$ & $\bm{\lambda}$ \\
			\midrule
			\mobse1\_D & \mobse{}1 & delayed & 1.0 & 0.1\\
			\mobse1\_R & \mobse{}1 & rapid & 1.0 & 0.1\\
			\mobse2\_D & \mobse{}2 & delayed & 1.0 & 0.1 \\
			\mobse2\_R & \mobse{}2 & rapid & 1.0 & 0.1\\
			\mobse1\_D1.5 & \mobse{}1 & delayed & 3.0 & 0.5\\
			\mobse1\_D0.02 & \mobse{}1 & delayed & 0.2 & 0.1\\
			\mobse2\_D1.5 & \mobse{}2 & delayed & 3.0 & 0.5\\
			\mobse2\_D0.02 & \mobse{}2 & delayed & 0.2 & 0.1\\
			\bottomrule	
		\end{tabular}
	\end{center}
	{\small Column 1: simulation name; column 2: stellar wind model (\mobse1 and \mobse2 see sec.~\ref{sec:2.1}); column 3: SN model (delayed and rapid from \citealt{Fryer2012}); column 4 and 5: values of $\alpha$ and $\lambda$ in the CE formalism. 
	}
	
\end{table}

\section{Results}
\begin{figure}
	\centering		
	\includegraphics[scale=0.36]{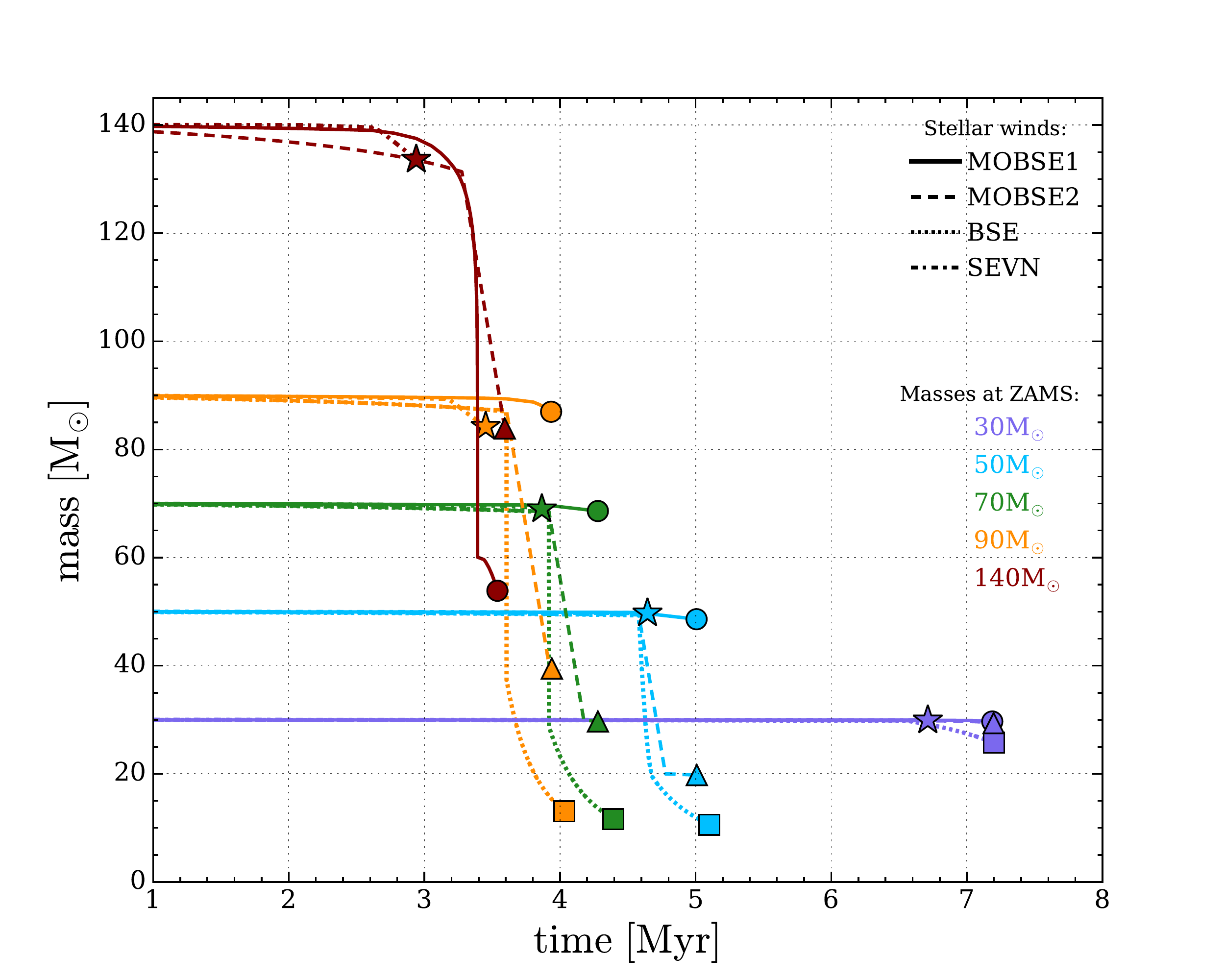}
	\caption{Stellar mass evolution with time for five different $M_{\mathrm{ZAMS}}$ at $Z=0.0002$ computed with \mobse, \bse{} and \sevn. Solid lines: \mobse1; dashed lines: \mobse2; dotted lines: \bse; dash-dot lines: \sevn. The markers identify the final mass of the stars: circles for \mobse1, triangles for \mobse2, squares for \bse~and stars for \sevn.}
	\label{fig:stellarwinds}
\end{figure}

\subsection{Mass loss by stellar winds}	
\label{sec:3.1}
In this Section, we discuss the evolution of single massive stars obtained with \mobse1 and \mobse2, in comparison to other open-source population synthesis codes, namely \bse{} \citep{Hurley2000,Hurley2002} and \sevn{} \citep{Spera2015,Spera2016,Spera2017}. \bse{} is the original code from which \mobse{} derives, while \sevn{} is a more recent code. In \bse{} single stellar evolution is implemented using polynomial fitting formulas \citep{Hurley2000}, while in \sevn{} stellar evolution is calculated from look-up tables (the current default tables are based on the recent {\sc PARSEC} stellar evolution code, \citealt{Bressan2012,Tang2014,Chen2015}).

Figure  \ref{fig:stellarwinds} shows the evolution of stellar mass at $Z=0.0002$
 for various ZAMS masses. At low ZAMS masses ($\lesssim 30$ \msun) the behavior of the considered codes is quite indistinguishable. The main difference is the duration of stellar life in \bse{} and \mobse{} with respect to the more updated \sevn{} code (see Fig.~\ref{fig:stellarwinds}). 
 
\begin{figure}
	\centering		
	\includegraphics[scale=0.36]{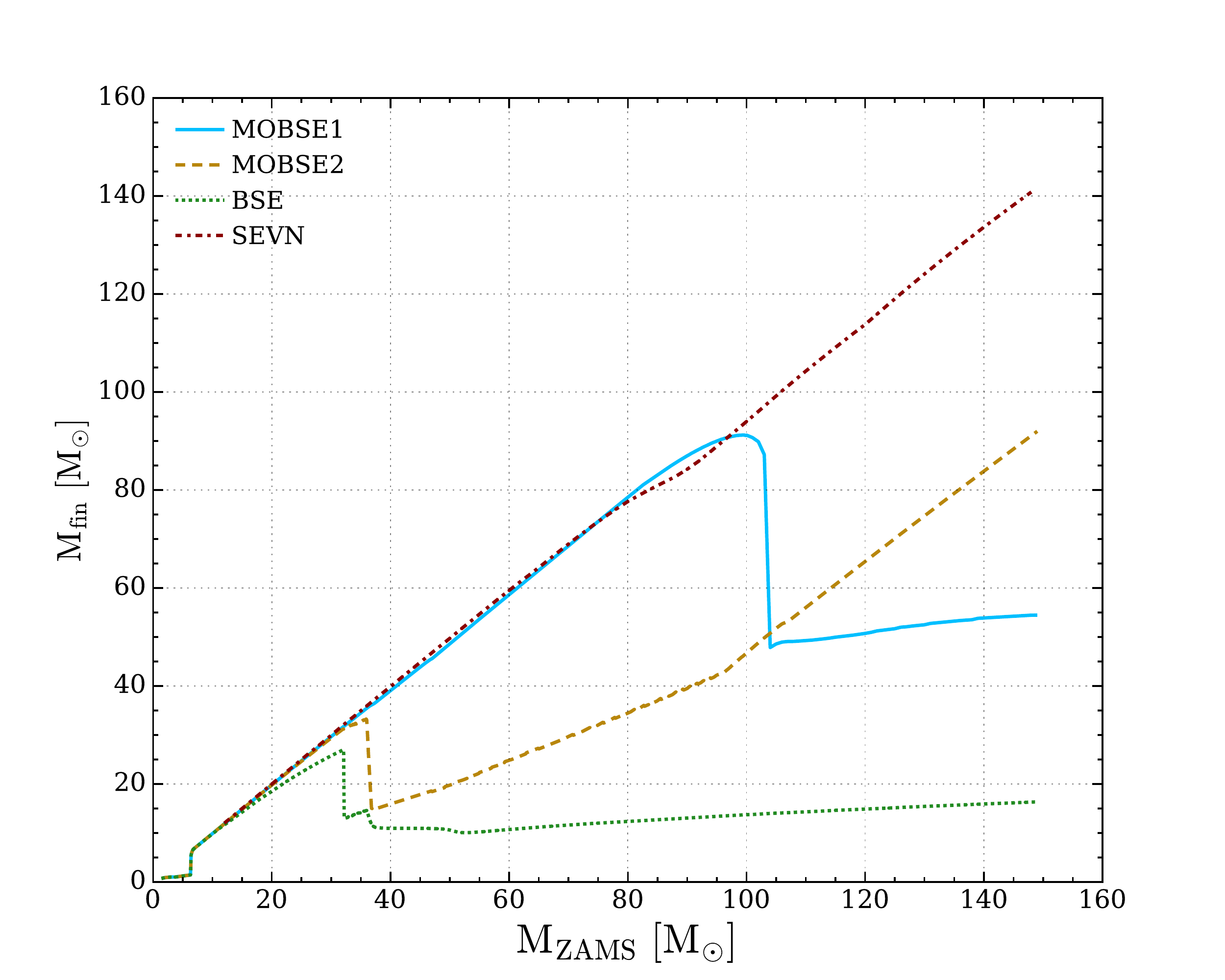}
	\caption{Final mass of a star ($M_{\mathrm{fin}}$) as a function of its ZAMS mass  at $Z = 0.0002$. Solid line: \mobse1; dashed line: \mobse2; dotted line: \bse; dash-dot line: \sevn. In all cases we considered the delayed SN model.\label{fig:mfin}}
\end{figure}

For larger ZAMS masses, there is a pronounced difference in the late evolutionary stages, due to the different stellar wind models. These differences are highlighted in Figure \ref{fig:mfin}, which shows the final stellar mass ($M_{\rm{fin}}$), as a function of the ZAMS mass ($M_{\rm{ZAMS}}$) at $Z=0.0002$. \mobse1 is in remarkable agreement with  \sevn{} for stars lighter than $M_{\rm{ZAMS}}\simeq 100 \msun$, predicting a low mass loss during the entire star's life. For more massive stars ($M_{\rm{ZAMS}}> 100 \msun$), \sevn{} follows the same trend as for lighter stars, i.e. mass loss is extremely quenched, while the value of $M_{\rm{fin}}$ in model \mobse1  drops to $\sim 50 \msun$.

The evolution of the stellar mass in \mobse2 is generally intermediate between that of \bse{} and that of both \sevn{} and \mobse1. This difference arises mainly from the treatment of LBV stars. In both \mobse1 and \sevn{} the mass loss scales as $\dot{M}\propto{}Z^{0.85}$, unless a star is radiation-pressure dominated ($\Gamma{}_e>2/3$, see equations~\ref{eq:scaling}). In contrast, \mobse2 assumes a strong mass loss rate for LBVs, independent of metallicity even if $\Gamma{}_e<2/3$ (see equation~\ref{eq:scaling2}).

\subsection{Mass spectrum of compact remnants}
\label{sec:3.2}

\begin{figure*}
	\includegraphics[scale=0.31]{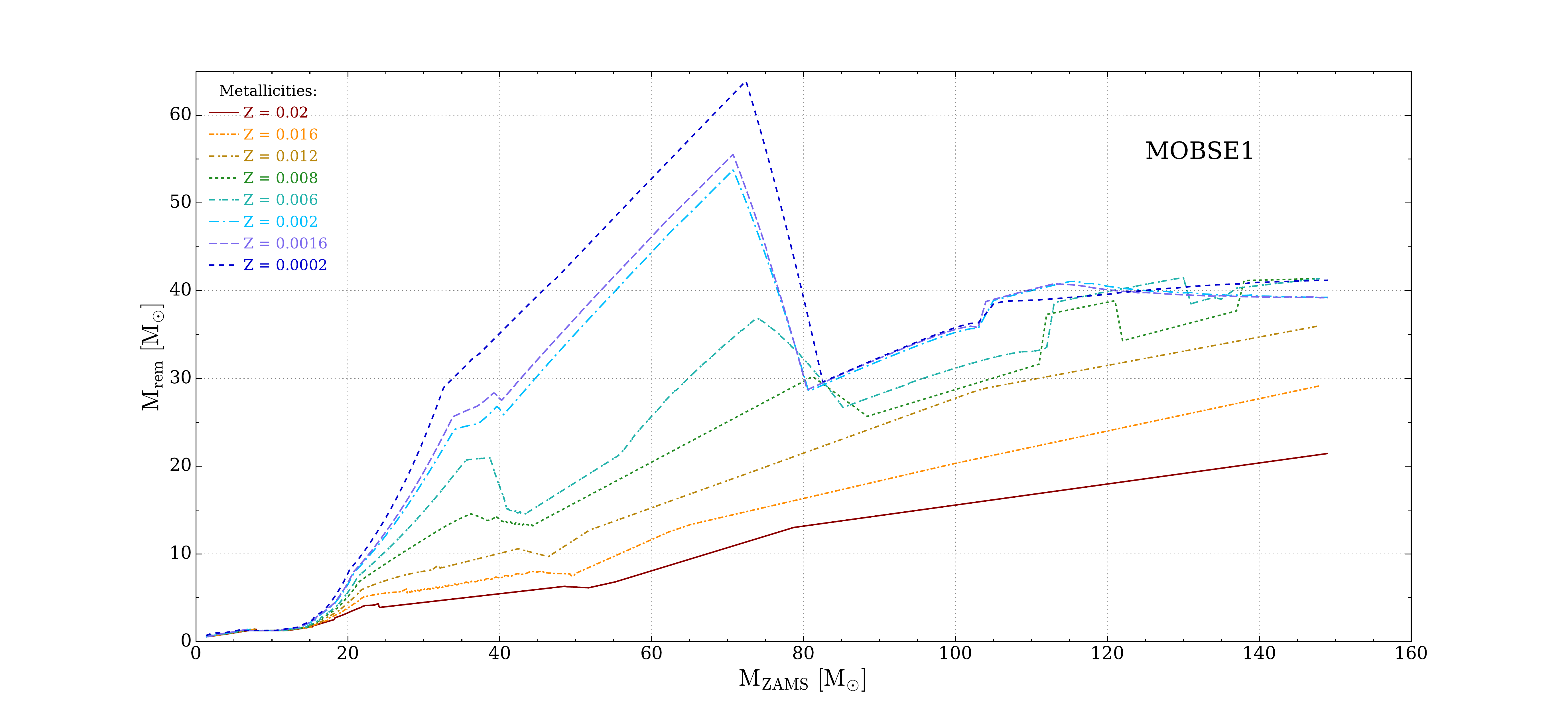}
	\includegraphics[scale=0.31]{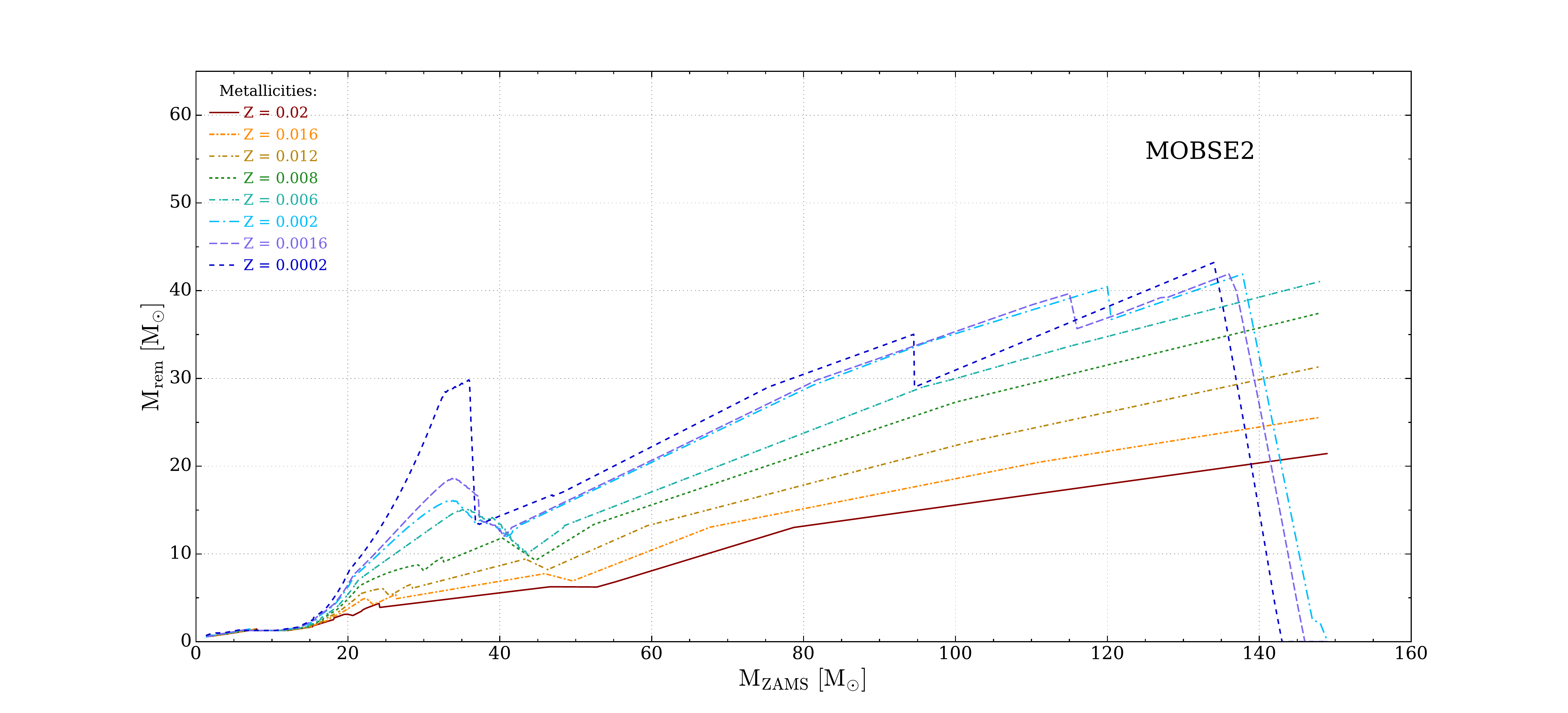}
	\caption{Mass of the compact remnant ($M_{\mathrm{rem}}$) as a function of the ZAMS mass  of the progenitor star, for different metallicities between $Z=0.0002$ and  $0.02$. Top: \mobse1. Bottom: \mobse2. In both cases we assume the delayed SN model.}
	\label{fig:massspec}
\end{figure*}	

Figure \ref{fig:massspec} shows the trend of the remnant mass ($M_{\rm{rem}}$) as a function of $M_{\rm ZAMS}$, for different values of the metallicity ($0.0002\leq Z \leq 0.02$). In this Figure, we adopt the delayed mechanism for SN explosions. The upper and the lower panel show the results of \mobse1 and \mobse2, respectively. As expected, there is a relation between the maximum mass of the compact remnants and the metallicity: the lower the metallicity is, the higher the mass of the heaviest remnant. 

At high metallicity ($Z=0.02$)  \mobse1 and \mobse2 are almost indistinguishable. In both models, the remnant mass increases monotonically with the ZAMS mass, until it reaches $M_{\rm{rem}} \sim 20 \msun$.

At lower metallicities ($Z= 0.002$ and $Z= 0.0002$), \mobse1 and \mobse2 differ significantly, especially for  $M_{\rm{ZAMS}} > 30\msun$. In \mobse1 the remnant mass rapidly increases with $M_{\rm ZAMS}$ untill it reaches its maximum ($M_{\rm rem}\sim{}50-65$ M$_\odot$) at $M_{\rm{ZAMS}} \sim 70 \msun$. For larger ZAMS masses, the remnant mass drops to $\sim{}40$ M$_\odot$. In \mobse2 the remnant mass has a local maximum for  $M_{\rm{ZAMS}} \sim 30 \msun$. For larger ZAMS masses, it drops and then rises steadily to a global maximum of $M_{\rm{rem}} \sim 40 \msun$ at $M_{\rm ZAMS}>120$ M$_\odot$.

Thus, the maximum BH mass predicted by \mobse1 is $50-65$ M$_\odot$, significantly higher than the maximum BH mass predicted by \mobse2 ($\sim{}40$ M$_\odot$). The main reason for this difference is, again, the dependence of mass loss on the Eddington factor implemented in \mobse1 but not in \mobse2.

\begin{figure}
	\includegraphics[scale=0.36]{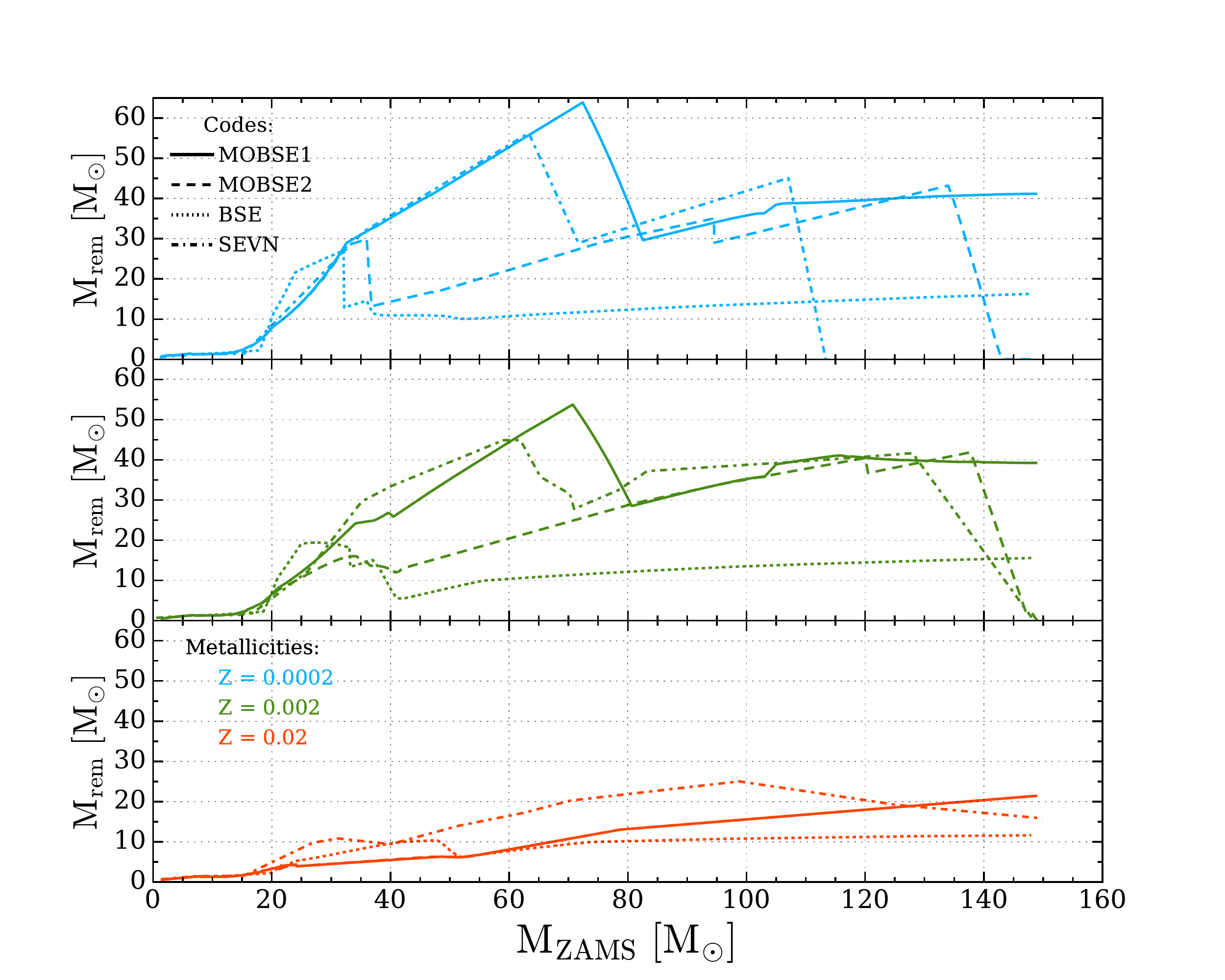}
	\caption{Mass of the compact remnant ($M_{\mathrm{rem}}$) as a function of the ZAMS mass of the progenitor stars, for $Z = 0.02$ (bottom panel), $Z = 0.002$ (central panel) and $Z = 0.0002$ (top panel). Solid lines: \mobse1; dashed lines: \mobse2; dash-dot lines: \sevn{};  dotted lines: \bse{}. In all cases the delayed SN model is assumed.}
	\label{fig:massspec4codes}
\end{figure}

Figure \ref{fig:massspec4codes} compares the mass spectrum of compact remnants obtained with \mobse{}, \sevn{} and \bse{} at three different metallicities. The dependence on metallicity is weaker for  \bse{}, and stronger for both \mobse{} and \sevn{}. At solar metallicity, all codes predict remnant masses $<30$ M$_\odot$. At metallicity $Z=0.002$ and $Z=0.0002$, \bse{} is the only code  predicting $M_{\rm rem}<30 $ M$_\odot$, because of the stellar wind prescriptions. Both \mobse{} and \sevn{} predict larger masses. We note that the BH mass spectrum obtained with \mobse2 is in agreement with \citet{Belczynski2016}, who adopt similar prescriptions for the stellar winds \citep{Belczynski2010}.

At low metallicity ($Z<0.002$)  \sevn{} and \mobse1 are in good agreement, while the remnant masses obtained with \mobse2 are significantly different from both \sevn{} and \mobse1. The key ingredient is again the fact that the stellar wind prescriptions implemented in  \sevn{} and \mobse1 depend on both the metallicity and the Eddington factor.

The good agreement between \sevn{} and \mobse1 is particularly remarkable, because \sevn{} \citep{Spera2017} adopts very recent stellar evolution models (from {\sc PARSEC}, \citealt{Bressan2012,Chen2015}), while \mobse1 is still based on the polynomial fitting formulas described by \cite{Hurley2000}. This result confirms the importance of accounting for the Eddington factor in mass loss models.

\begin{figure}
	\includegraphics[scale=0.36]{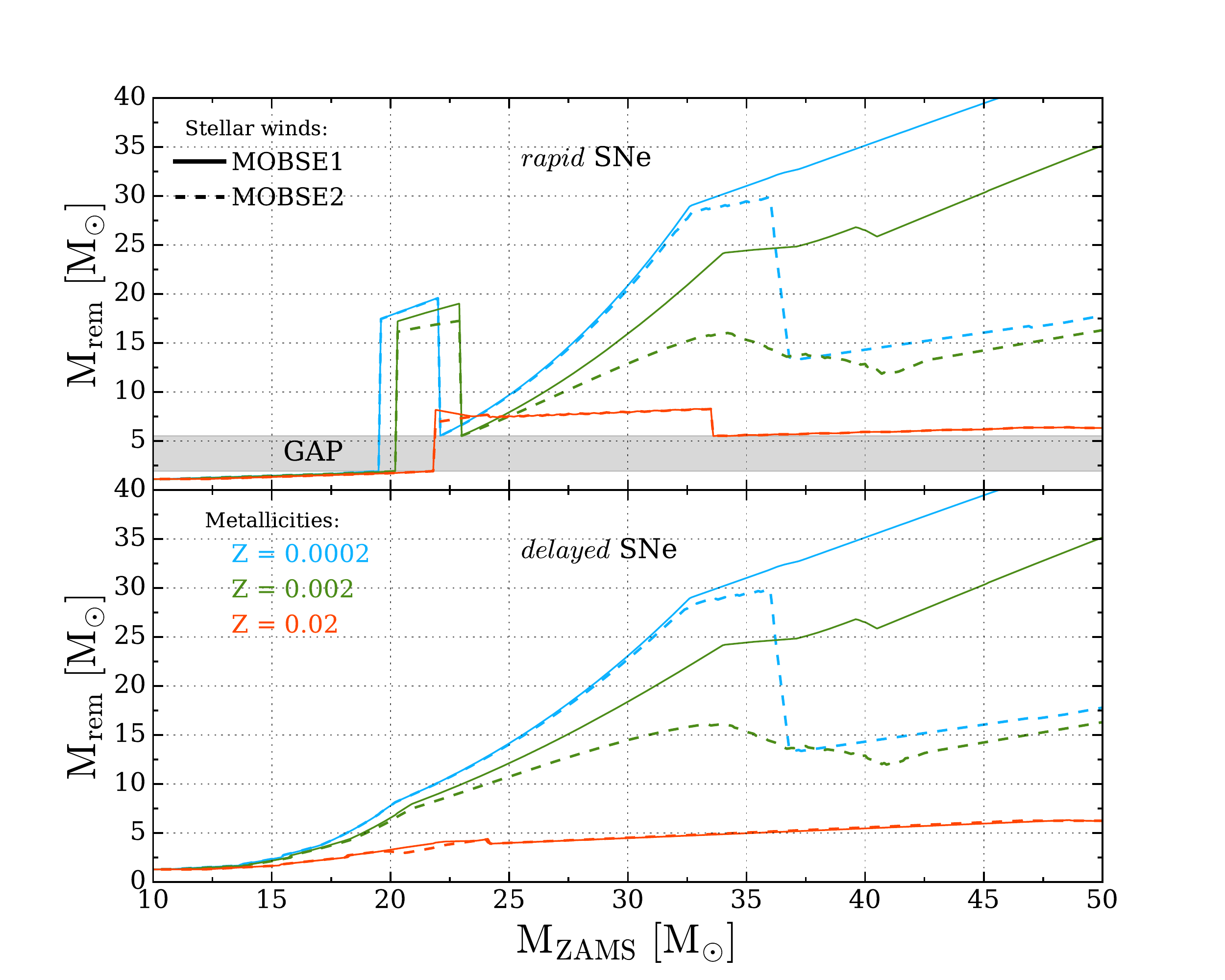}
	\caption{Mass of the compact remnant ($M_{\mathrm{rem}}$) as a function of the ZAMS mass of the progenitor star ($M_{\mathrm{ZAMS}}$) for two different core-collapse SN models: rapid SN model (upper panel) and delayed SN model (bottom panel). Red lines: metallicity $Z = 0.02$; green lines: $Z = 0.002$; blue lines: $Z = 0.0002$. Solid lines: \mobse1, dashed lines: \mobse2. In the top panel the mass gap between the heaviest NSs and the lightest BHs ($\sim 2 \mathrm{M}_{\odot}$ to $\sim 5 \msun$) is highlighted by a shaded area.}
	\label{fig:massspeczoom}
\end{figure}	

In Figure~\ref{fig:massspec4codes} we can also note the effect of PISNe and PPISNe on the most massive metal-poor stars (see also Figure~\ref{fig:PPISN} in Appendix~\ref{sec:appB}). At metallicity $Z\leq{}0.002$, very massive stars leave no remnant as an effect of PISNe in both \mobse2 and \sevn. PISNe do not occur in \mobse1 and \bse, because the final Helium core mass is always below the threshold for PISNe ($\sim{}64$ M$_\odot$). PPISNe occur in \mobse1, \mobse2 and \sevn{} at low metallicity ($Z\leq{}0.008$ for \mobse1 and \sevn{} and $Z\leq{}0.002$ for \mobse2) for stars with $32\,{}{\rm M}_\odot\lesssim{}M_{\rm He,f}\lesssim{}64\,{}{\rm M}_\odot$. Their effect is a substantial decrease of the remnant mass with respect to the final mass of the star ($M_{\rm fin}$, Figure~\ref{fig:mfin}). PPISNe do not occur in \bse{}, because the final Helium core mass does not reach the threshold for PPISNe in \bse{}.


Figure~\ref{fig:massspeczoom} compares the rapid and the delayed core-collapse SN models. ZAMS masses larger than 50 M$_\odot$ are not shown, because the rapid and the delayed SN models produce exactly the same remnant mass for $M_{\rm ZAMS}>50$ M$_\odot$, in agreement with \cite{Fryer2012} and \cite{Spera2015}.

The main difference between rapid and delayed SN model is the number of remnants with mass $2<M_{\rm rem}<5$ M$_\odot$. The rapid SN model predicts a mass gap between the lightest BHs ($\sim 5\msun$) and the heaviest NSs ($\sim 2\msun$), while the delayed model predicts no gap. This result is consistent with previous studies \citep{Fryer2012,Spera2015}. Dynamical mass measurements of compact objects in X-ray binaries show  marginal indications for the existence of a mass gap (e.g. \citealt{Ozel2010,Farr2011}), possibly suggesting a preference for a rapid SN explosion.

\subsection{Black hole binaries (BHBs)}
\label{sec:4}
\begin{figure*}
	\includegraphics[width=0.85\textwidth]{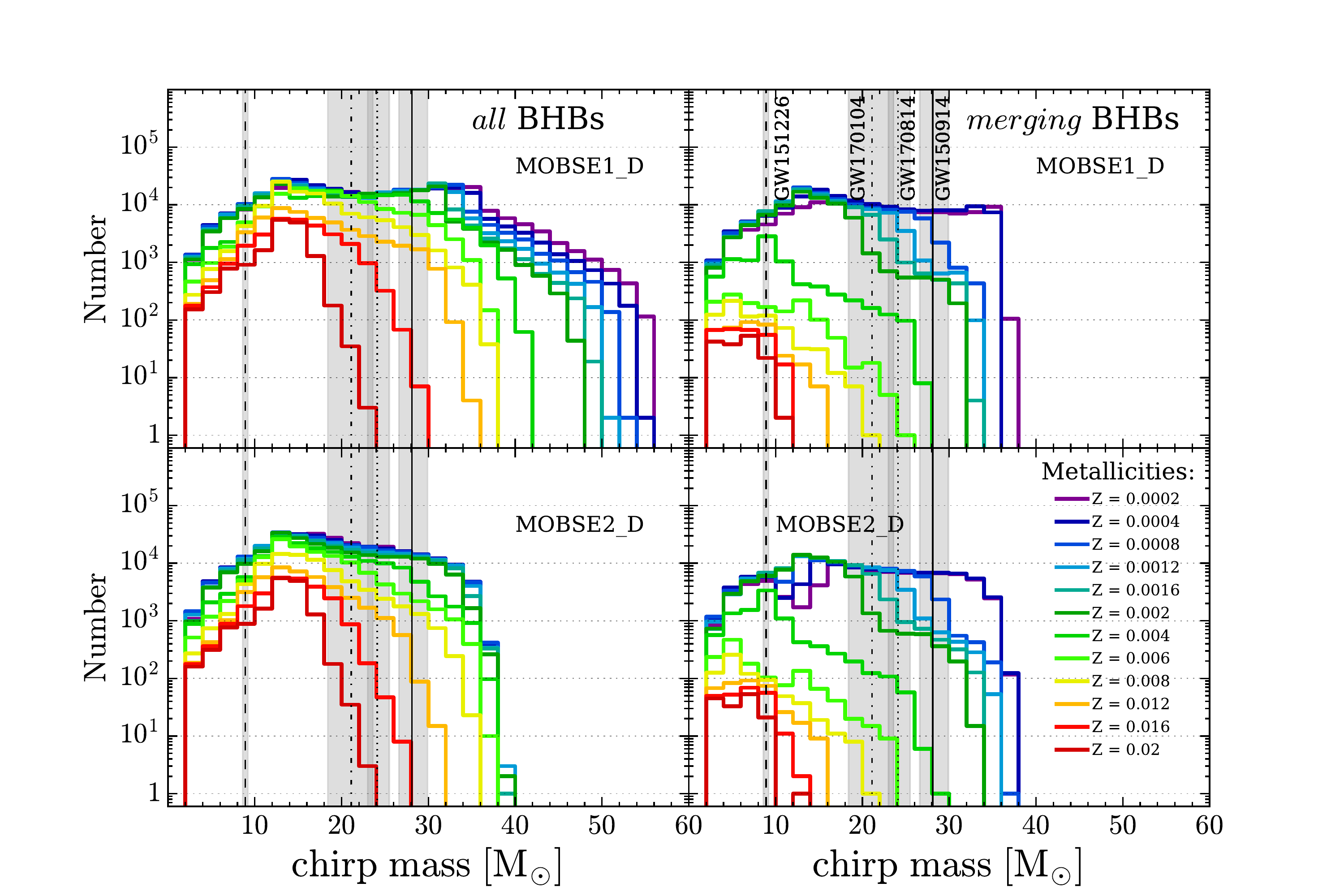}
	
	\caption{Chirp mass distribution of BHBs for \mobse{}1\_D (upper panels) and \mobse{}2\_D (lower panels). Left-hand column: chirp mass distribution for all BHBs. Right-hand column: chirp mass distribution for the merging BHBs only.  Solid lines represent the chirp mass distributions at different metallicity, ranging from 0.0002 to 0.02. The vertical lines in all panels are the chirp masses of GW151226, GW150914 \citep{Abbott2016a}, GW170104 \citep{Abbott2017} and GW170814 \citep{LIGO2017} with the corresponding uncertainties ( at the 90 per cent credible level, shadowed regions). \label{fig:mchirp}}
\end{figure*}
\begin{figure*}
		\includegraphics[width=0.85\textwidth]{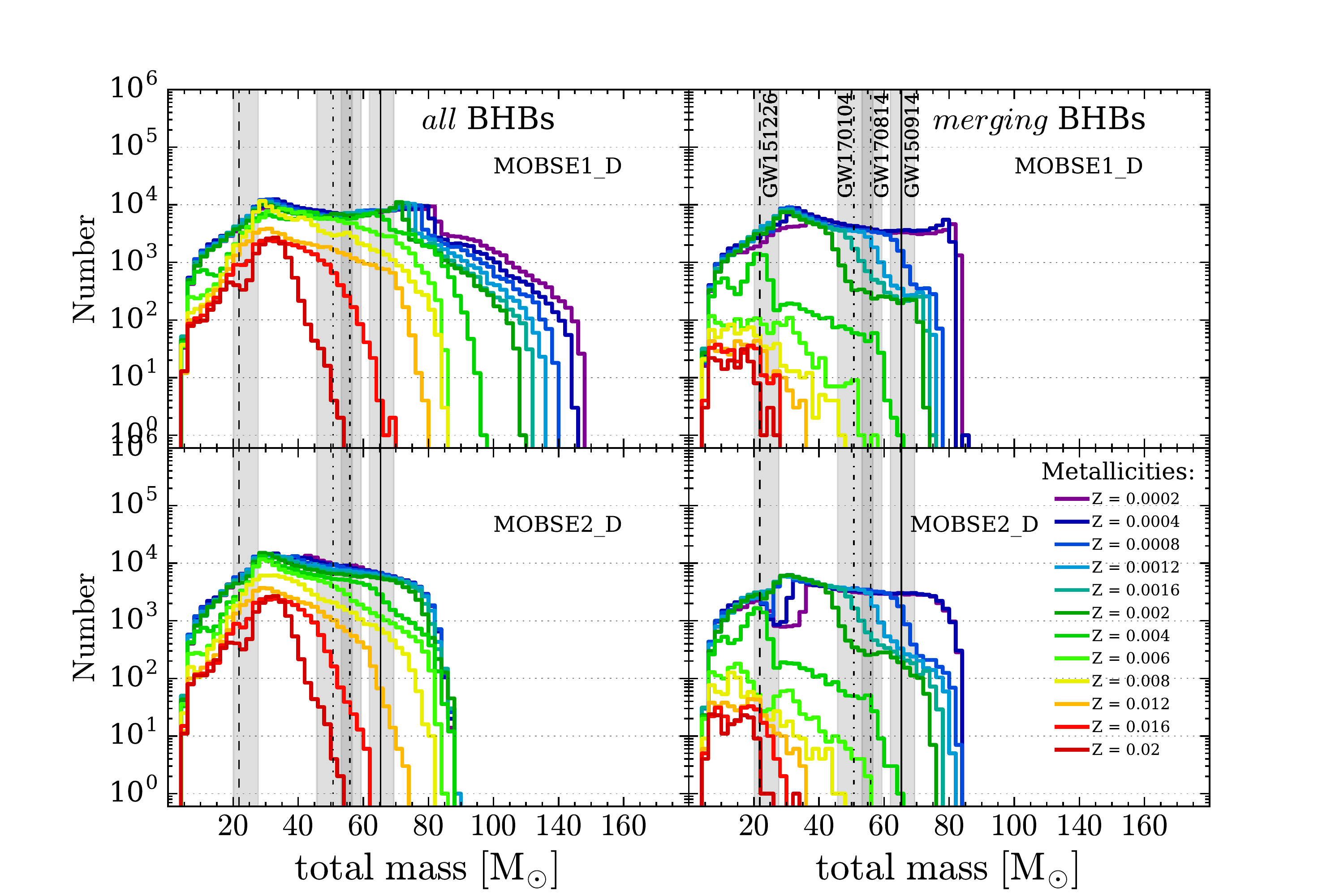}
	\caption{Same as Figure~\ref{fig:mchirp}, but for the distribution of total masses.\label{fig:mtot}}

\end{figure*}

\begin{figure*}
	\includegraphics[width=1.05\textwidth]{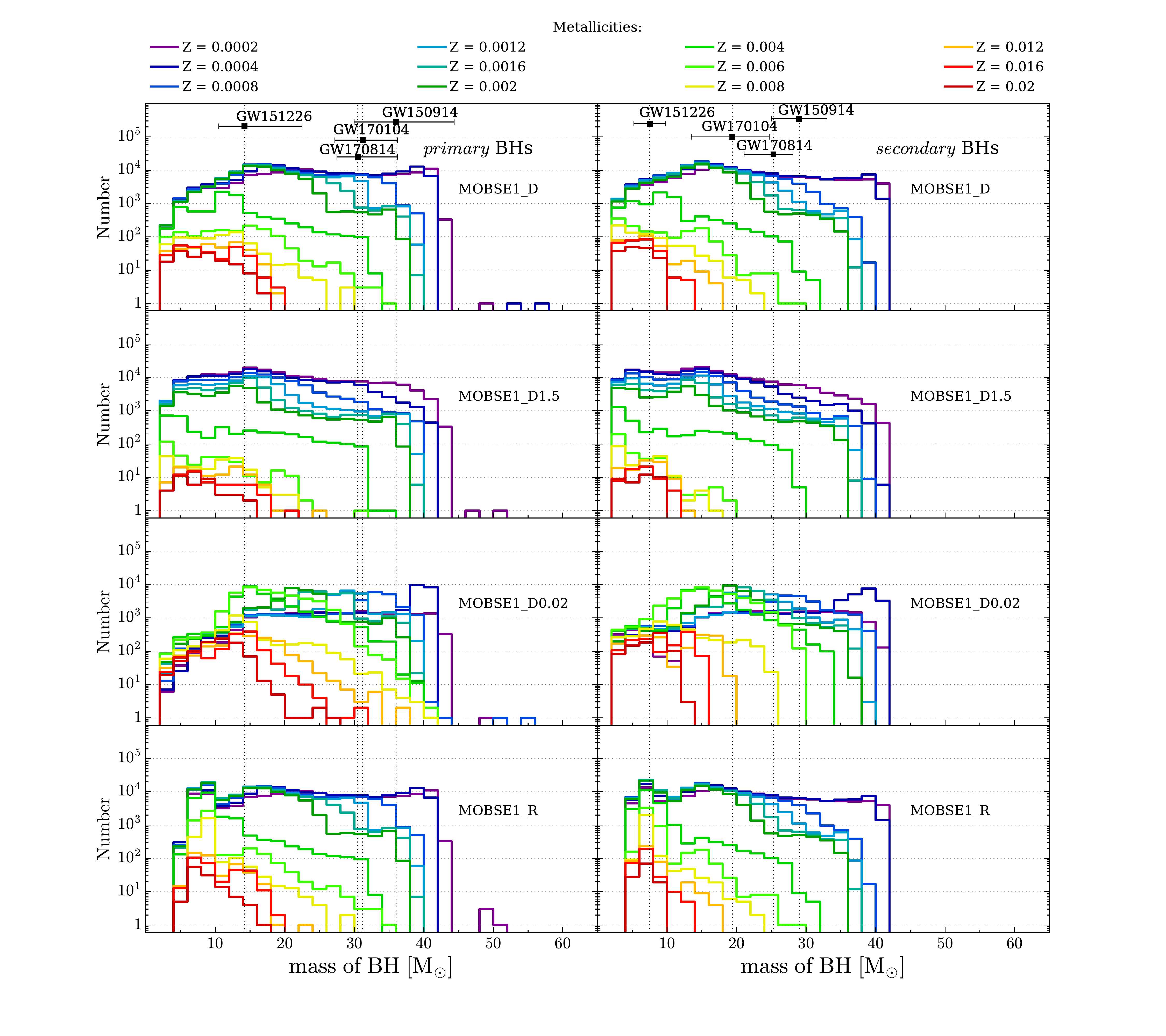}
	\vspace*{-10mm}
	\caption{Mass distribution of primary (left-hand column) and secondary members (right-hand column) of merging BHBs in the eight simulation sets. Line colours (from red to violet) correspond to decreasing metallicity (from $Z=0.02$ to $Z=0.0002$). From top to bottom: \mobse{}1\_D, \mobse{}1\_D1.5, \mobse{}1\_D0.02, and \mobse{}1\_R. See Table~\ref{tab:simulations} for details on each simulation. The vertical lines on the left-hand (right-hand) column are the mass of the primary (secondary) BH in GW151226,  GW150914 \citep{Abbott2016a}, GW170104 \citep{Abbott2017} and GW170814 \citep{LIGO2017}. The error bars show the uncertainties on each mass ( at the 90 per cent credible level). \label{fig:Cmergingmbh}}
\end{figure*}	

\begin{figure*}
	\includegraphics[width=1.05\textwidth]{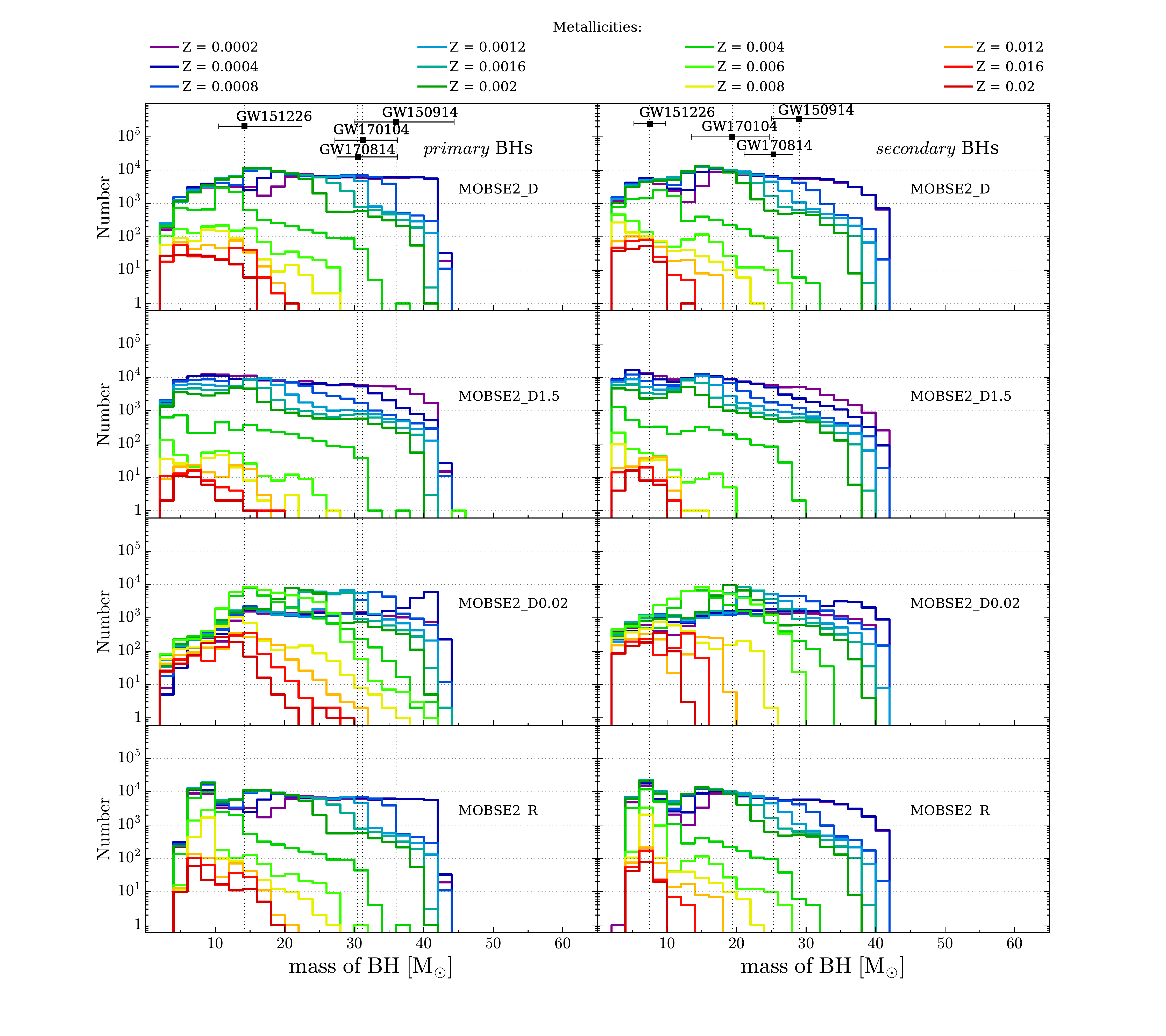}
	\vspace*{-10mm}
	\caption{Same as Fig.~\ref{fig:Cmergingmbh} but for simulation with \mobse2. From top to bottom: \mobse{}2\_D, \mobse{}2\_D1.5, \mobse{}2\_D0.02, and \mobse{}2\_R.  \label{fig:Bmergingmbh}}
\end{figure*}

\begin{figure*}
	\includegraphics[width=1.05\textwidth]{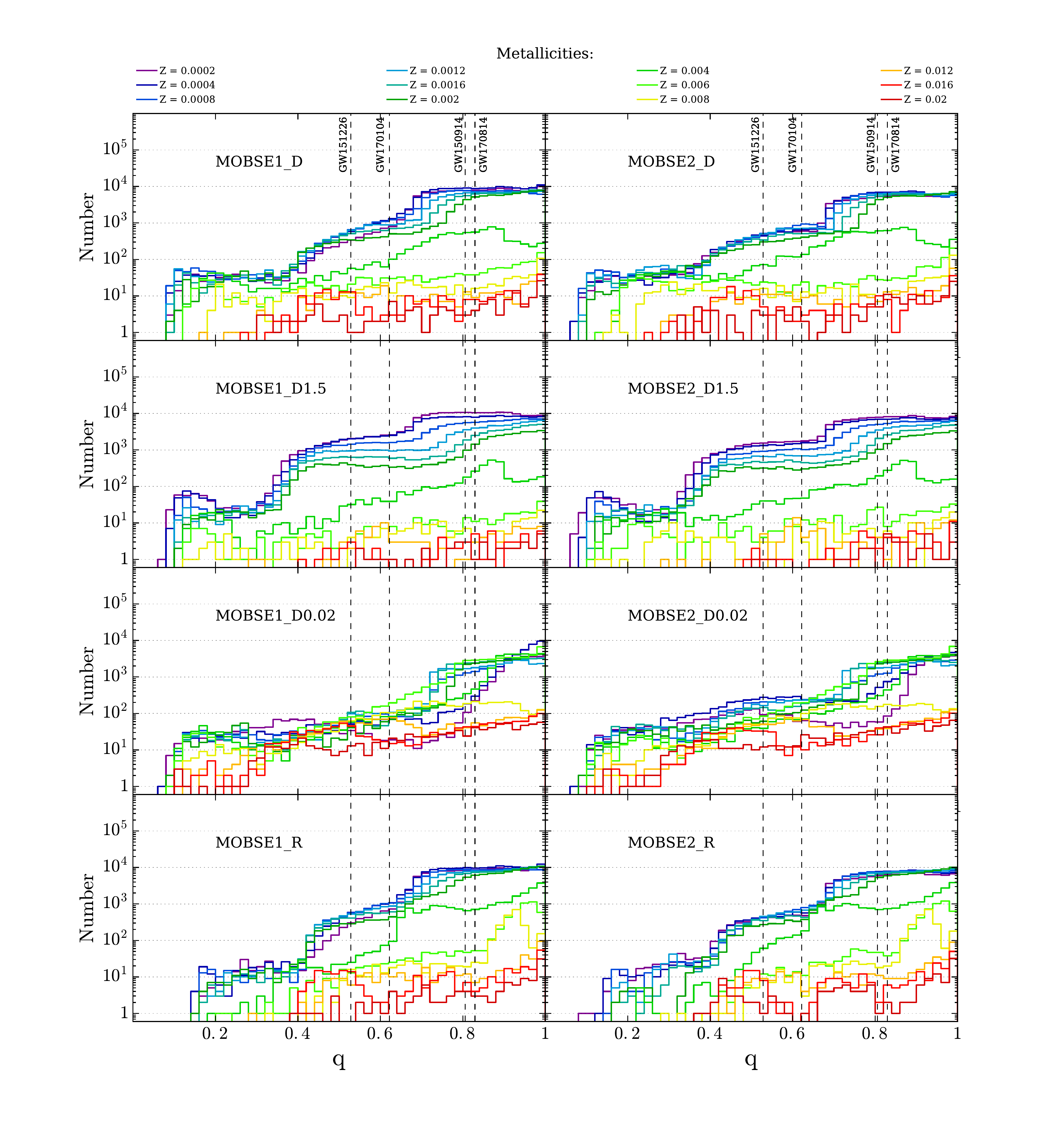}
	\vspace*{-10mm}
	\caption{Distribution of the mass ratio $q=m_{\rm s}/m_{\rm p}$ (where $m_{\rm p}$ and $m_{\rm s}$ are the mass of the primary and of the secondary BH, respectively) of merging BHBs. Left-hand panel, from top to bottom: \mobse{}1\_D, \mobse{}1\_D1.5, \mobse{}1\_D0.02, \mobse{}1\_R. Right-hand panel, from top to bottom: \mobse{}2\_D, \mobse{}2\_D1.5, \mobse{}2\_D0.02, \mobse{}2\_R. The vertical lines are the mass ratio of GW151226, GW150914 \citep{Abbott2016a}, GW170104 \citep{Abbott2017} and GW170814 \citep{LIGO2017}. \label{fig:massratio}}
\end{figure*}	

 In this section, we focus on the properties of BHBs derived from our binary population-synthesis simulations with \mobse{}. The left-hand panels of Figures \ref{fig:mchirp} and \ref{fig:mtot} show the chirp mass\footnote{ The chirp mass is named that because it is this combination of $m_{\rm p}$ and $m_{\rm s}$ that determines how fast the binary sweeps, or chirps, through a frequency band. In fact, it can be shown that the amplitude and the frequency of GWs scale as $m_{\rm chirp}^{5/3}$ and $m_{\rm chirp}^{-5/8}$, respectively \citep{Maggiore2008}.} ($	m_{\rm{chirp}} = (m_{\rm p}\,{}m_{\rm s})^{3/5} / (m_{\rm p} + m_{\rm s})^{1/5}$, where $m_{\rm p}$ and $m_{\rm s}$ are the masses of the primary and secondary BH, respectively) and the total mass distributions of all BHBs which formed in our simulations \mobse{}1\_D and \mobse{}2\_D, including both BHBs which merge and BHBs which do not merge in a Hubble time. Simulations labelled as \mobse{}1\_D (\mobse{}2\_D) were run with \mobse{}1 (\mobse{}2), adopting the delayed SN model and assuming $\alpha=1$, $\lambda=0.1$ for the CE (see Table \ref{tab:simulations} for details on the simulations). 

At low metallicity ($Z\lesssim{}0.0004$), \mobse{}1 produces very massive BHBs, with total mass up to $M_{\rm{tot}} \simeq 150 \msun$ and chirp mass up to $M_{\rm{chirp}} \simeq 55 \msun$, while the heaviest BHBs obtained with \mobse{}2  have $M_{\rm{tot}} \simeq 90 \msun$ and $M_{\rm{chirp}} \simeq 40 \msun$. 
At solar metallicity ($Z=0.02$), the maximum chirp mass (total mass) is $M_{\rm{chirp}} \simeq 20 \msun$ ($M_{\rm{tot}} \simeq 50 \msun$) for both \mobse{}1 and \mobse{}2. 


We now restrict our attention to merging BHBs. The right-hand panels of  Figures \ref{fig:mchirp} and \ref{fig:mtot} show the chirp mass and the total mass distributions for the sub-sample of merging BHBs (defined as BHBs which merge within a Hubble time). It is apparent that the maximum mass of merging BHBs is significantly smaller than the maximum mass of non-merging BHBs. This difference persists at all metallicities, and is more pronounced in \mobse1 than in \mobse2. 

The heaviest merging BHs have $M_{\rm{BH,max}} \simeq 45 \msun$ (only few systems have a primary BH $\gtrsim{}55$ M$_\odot$, in model \mobse1) at $Z\leq 0.0002$ and $M_{\rm{BH,max}} \simeq 20 \msun$ at $Z=0.02$. The maximum values of the chirp masses are $M_{\rm{chirp}} \simeq 40 \msun$ and $M_{\rm{chirp}} \simeq 10 \msun$ at $Z=0.0002$ and $Z=0.02$, respectively. The masses of merging BHBs predicted by \mobse1 are remarkably similar to those predicted by \mobse2. This fundamental difference between merging BHBs and other BHBs holds for all eight sets of simulations performed in this paper, included those which are not shown in Figures~\ref{fig:mchirp} and \ref{fig:mtot}. This happens because the most massive BHBs in \mobse{}1 come from massive ($\sim{}60-80$ M$_\odot$) progenitors which die as red super-giant stars. Thus, all such BHBs form with very large semi-major axis (otherwise their progenitors merge prematurely, due to their large radii) and cannot merge within a Hubble time.

\begin{table*}
	\begin{center}
		\caption{Minimum metallicity of progenitors of GW events.\label{tab:Z}}
		\begin{tabular}{lcccccccc}
			\toprule
	& \multicolumn{4}{c}{\mobse{}1} & \multicolumn{4}{c}{\mobse{}2} \\ \cmidrule(lr){2-5}\cmidrule(lr){6-9}
 GW event  & D1.5 & D  & D0.02 & R & D1.5 & D & D0.02 & R \\
			\midrule
 GW150914 & $Z\leq 0.002$ & $Z\leq 0.004$ & $Z\leq 0.006$ & $Z\leq 0.004$ & $Z\leq 0.002$ & $Z\leq 0.004$ & $Z\leq 0.006$ & $Z\leq 0.004$ \\
LVT151012 & $Z\leq 0.008$ & $Z\leq 0.012$ & $Z\leq 0.016$ & $Z\leq 0.012$ & $Z\leq 0.006$ & $Z\leq 0.012$ & $Z\leq 0.016$ & $Z\leq 0.012$ \\			
GW151226 & $Z\leq 0.02$  & $Z\leq 0.02 $ & $Z\leq 0.02$  & $Z\leq 0.02$  & $Z\leq 0.02$  & $Z\leq 0.02$  & $Z\leq 0.02$  & $Z\leq 0.02$  \\
GW170104 & $Z\leq 0.006$ & $Z\leq 0.006$ & $Z\leq 0.012$ & $Z\leq 0.006$ & $Z\leq 0.004$ & $Z\leq 0.006$ & $Z\leq 0.008$ & $Z\leq 0.006$ \\
GW170814 & $Z\leq 0.004$ & $Z\leq 0.006$ & $Z\leq 0.008$ & $Z\leq 0.006$ & $Z\leq 0.004$ & $Z\leq 0.004$ & $Z\leq 0.008$ & $Z\leq 0.004$ \\
\bottomrule	
		\end{tabular}
	\end{center}
\begin{flushleft}
	{\small Column 1: GW detection; column 2-9: maximum star metallicity at which we can obtain merging BHBs with the same mass as the detected ones in runs \mobse{}1\_D1.5, \mobse{}1\_D, \mobse{}1\_D0.02, \mobse{}1\_R, \mobse{}2\_D1.5, \mobse{}2\_D, \mobse{}2\_D0.02, and \mobse{}2\_R (see Table \ref{tab:simulations}).
	}
\end{flushleft}

\end{table*}

Figures \ref{fig:mchirp} to \ref{fig:mtot} also show that the number of BHBs scales inversely with the metallicity of the progenitors. This trend is particularly strong if we consider only merging BHBs. 
This result originates from several factors. At higher metallicity, stars radii are larger, and thus a larger fraction of binaries merge before becoming a BHB. Moreover, we assume stronger SN kicks for lower BH masses. Thus, SN kicks are more efficient in unbinding light binaries, which are more common at high metallicity.

Figure~\ref{fig:Cmergingmbh} shows the distribution of masses of the primary BH (i.e. the most massive one) and of the secondary BH (i.e. the least massive one) for all merging BHBs in the four runs with \mobse{}1. Figure~\ref{fig:Bmergingmbh} is the same for the four runs with \mobse{}2. There are no significant differences between merging BHB masses in \mobse{}1 and \mobse{}2, regardless of the SN model or CE prescription. 
The maximum mass of merging BHBs does not seem to depend significantly on the assumed SN model or on the assumed CE parameters. The minimum mass of merging BHBs does depend on the assumed SN model, because the rapid SN model (\mobse{}1\_R and \mobse{}2\_R) does not allow to form compact remnants with mass in the range of $2-5$ M$_\odot$. 

The main difference between different values of the CE parameters is the number of merging BHBs with relatively high metallicity ($0.006\leq{}Z\leq{}0.02$). The two models \mobse{}1\_D0.02 and \mobse{}2\_D0.02, which adopt $\alpha{}\,{}\lambda{}=0.02$, form a significantly larger number of merging BHBs with relatively high metallicity  ($0.006\leq{}Z\leq{}0.02$) than models with a larger value of $\alpha{}\,{}\lambda{}$. 
As already discussed in \cite{Mapelli2017}, this is likely due to the fact that a lower value of $\alpha{}\,{}\lambda{}$ makes the spiral in of the cores much more efficient, bringing the two cores on a much smaller final orbital separation. Thus, even binaries with a very large initial orbital separation might give birth to a merging BHB system, provided that they can enter a CE phase. Entering a CE phase is much easier for metal-rich stars, because their radii are larger than those of their metal-poor analogues.

 On the other hand, \mobse{}1\_D0.02 and \mobse2{}\_D0.02 also form a significantly smaller number of {\it light} merging BHBs ($<12$ M$_\odot$) with relatively low metallicity ($Z<0.006$) than models with a larger value of $\alpha{}\,{}\lambda{}$. This can be explained as follows.  
Assuming a lower value of $\alpha \lambda$ means that it is harder to eject the CE during the CE phase. This implies that the minimum semi-major axis ($a_{\rm{min}}$) above which a binary system survives the CE without merging is larger for a lower value of $\alpha \lambda$.

In the case of (both metal-rich and metal-poor) massive stars, the maximum stellar radii are always $\geq a_{\rm{min}}$ for the considered range of $\alpha \lambda$.  This means that changes of $\alpha \lambda$ (and consequently of  $a_{\rm{min}}$) do not affect the number of massive binary systems which merge prematurely (before becoming BHBs). In contrast, for light ($\lesssim 30\msun$) meta-poor stars ($Z\leq 0.006$) this difference in $a_{\rm{min}}$ is crucial, because the maximum stellar radii are $< a_{\rm{min}}$ for $\alpha \lambda = 0.02$ but $> a_{\rm{min}}$ for $\alpha \lambda \geq 0.1$. Therefore, the same binary will not survive the CE phase in the case with $\alpha \lambda = 0.02$, while it will survive without merging prematurely in the case with $\alpha \lambda \geq 0.1$.
This effect explains the dearth of merging BHBs with $M < 12 \msun$ in \mobse1\_D0.02 and \mobse2\_D0.02 simulations respect to \mobse1\_D,  \mobse2\_D, \mobse1\_D1.5 and \mobse2\_D1.5 simulations.

In addition, for metal-rich stars there is another effect that plays a role. Indeed, the spiral-in during CE is more efficient for small values of $\alpha \lambda$, so even initial larger binaries can become close binaries and eventually evolve into merging BHBs.

From Figures~\ref{fig:Cmergingmbh} and \ref{fig:Bmergingmbh} and from Table~\ref{tab:Z} it is apparent that our models can account for all four GW events reported so far. The most massive systems (GW150914, GW170104 and GW170814) can be generated only by metal-poor progenitors. In particular, GW150914-like systems are produced only for  $Z\leq 0.006$, GW170814-like systems for $Z\leq 0.008$, GW170104-like systems for $Z\leq 0.012$, and LVT151012-like systems for $Z\leq 0.016$, while GW151226-like systems exist at all metallicities ($Z\leq 0.02$, see Tab. \ref{tab:Z} for details). From Tab. \ref{tab:Z} it is also interesting to note that the higher $\alpha\lambda$ is, the lower the maximum metallicity to produce the observed GW events.

Finally, Figure~\ref{fig:massratio} shows the mass ratio between the secondary BH and the primary BH in the merging BHBs. While nearly equal-mass systems are more common in our models, we find merging BHBs with nearly all possible mass ratios, down to $q\sim{}0.1$. This is at odds with models of BHB formation through chemically homogeneous evolution \citep{Marchant2016,Demink2016,Mandel2016}, which predict the formation of nearly equal mass merging BHBs, but is  consistent with the mass ratio of the four GW detections.

\begin{figure}
	\centering
	\includegraphics[scale=0.34]{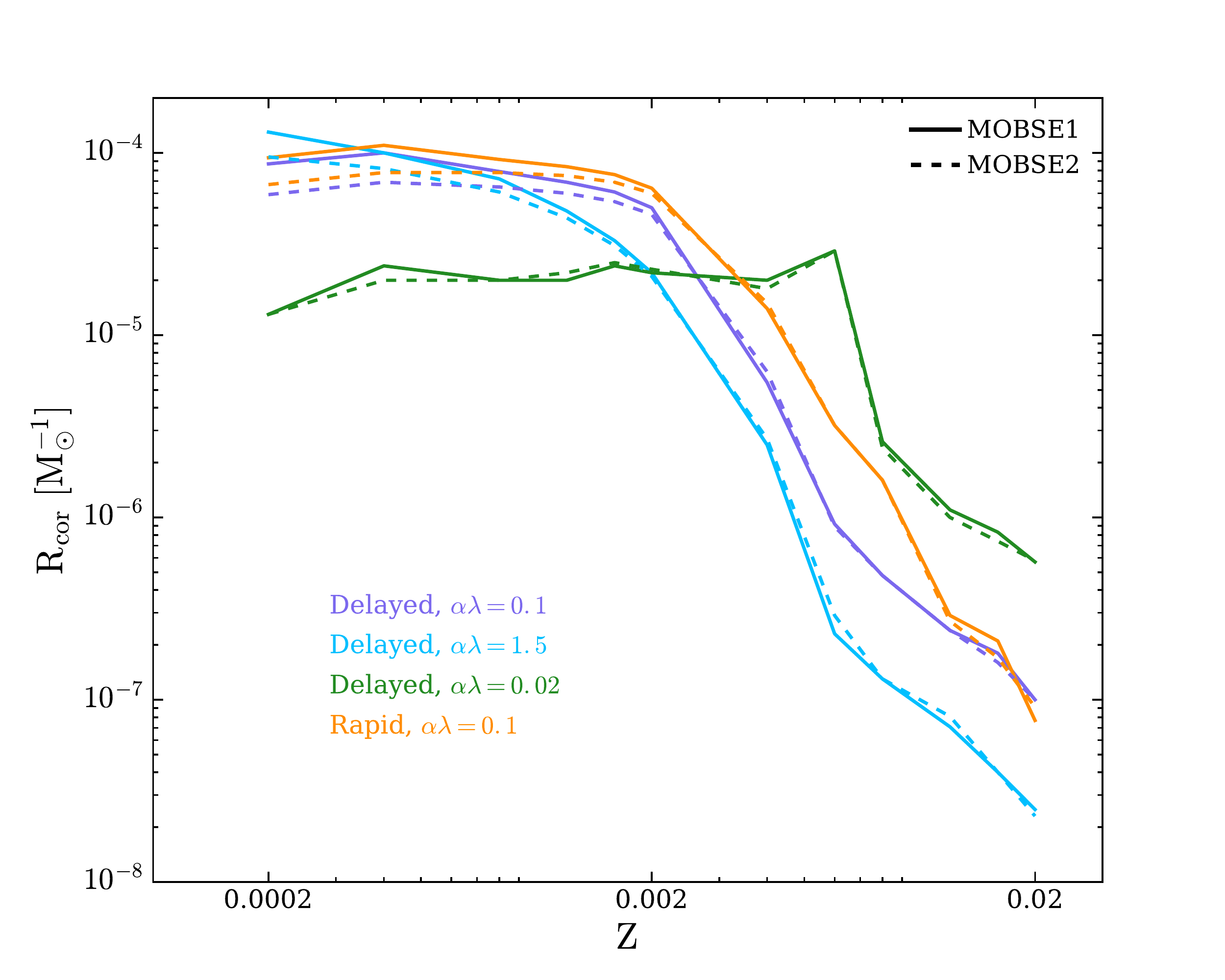}
	\caption{Corrected number of mergers per unit mass as a function of the metallicity for all sets of simulations. The colors identify different assumptions for the SN mechanism and for the values of CE parameters $\alpha$ and $\lambda$. Solid (dashed) lines: \mobse1 (\mobse2).    \label{fig:rate}}
\end{figure}

\subsection{Mergers per unit mass}
\label{sec:4.3}
For each simulation, we calculate the number of merging BHBs per unit mass as 
\begin{equation}
	R = \frac{N_{\mathrm{merger}}}{M_{\mathrm{tot,sim}}}~,
\end{equation}
where $N_{\mathrm{merger}}$ is the number of merging BHBs in each sub-sample and $M_{\mathrm{tot,sim}}$ is the total mass of the corresponding sub-sample.

This number must be corrected to account for the fact that we simulate only massive ($m_1>5$ M$_\odot$) binary systems (no single stars). We thus introduce the corrected number of merging BHBs per unit mass as
\begin{equation}
	R_{\rm{cor}} = f_{\rm  bin}\,{}\,{}f_{\rm IMF}\,{}\,{} R,
\end{equation}
where $f_{\rm bin}$ is the correction factor used to take into account that we only simulate binary systems. We put $f_{\rm bin}=0.5$, assuming that $50$ per cent of stars are in binaries \citep[see e.g.][]{Sana2013}. The correction factor $f_{\rm IMF}=0.285$ accounts for the fact that we actually simulate only systems with primary mass $m_1 \geq 5\msun$. The values of $R$ and $R_{\rm cor}$ are shown in Tables \ref{tab:mergersC} and \ref{tab:mergersB}, respectively.

Figure \ref{fig:rate} shows $R_{\rm{cor}}$ as a function of the metallicity $Z$ for all simulations. $R_{\rm{cor}}$ strongly depends on the metallicity. In particular, $R_{\rm{cor}}$ is $\gtrsim{}2$ orders of magnitude higher at low metallicity ($Z\lesssim{}0.002$) than at high metallicity ($Z\sim{}0.02$). This means that we form much more merging BHBs if the progenitors are metal-poor stars  (see Tables \ref{tab:mergersC} and \ref{tab:mergersB}). 

\begin{table*}
	\begin{center}
		\caption{ Merging BHBs in \mobse{}1.\label{tab:mergersC}}
		\begin{tabular}[t]{lcccccccc}
			\toprule
			 & \multicolumn{2}{c}{ \mobse{}1\_D1.5}  & \multicolumn{2}{c}{ \mobse{}1\_D} & \multicolumn{2}{c}{ \mobse{}1\_D0.02} & \multicolumn{2}{c}{ \mobse{}1\_R}\\
			 \cmidrule(lr){2-3}\cmidrule(lr){4-5}\cmidrule(lr){6-7}\cmidrule(lr){8-9}
		{Z}	& { N$_{\mathrm{merger}}$} & { R$_{\rm{cor}}$ [\msun$^{-1}$]} & { N$_{\mathrm{merger}}$} & { R$_{\rm{cor}}$ [\msun$^{-1}$]} & { N$_{\mathrm{merger}}$} & { R$_{\rm{cor}}$ [\msun$^{-1}$]} & { N$_{\mathrm{merger}}$} & { R$_{\rm{cor}}$ [\msun$^{-1}$]}\\
			\midrule 
0.0002	&189837 &1.3E-04        &131819 &8.7E-05        &20259  &1.3E-05        &142137 &9.4E-05\\
0.0004	&157034 &1.0E-04        &152054 &1.0E-04        &35744  &2.4E-05        &164478 &1.1E-04\\
0.0008	&108094 &7.2E-05        &119200 &7.9E-05        &29503  &2.0E-05        &139294 &9.2E-05\\
0.0012	&73138  &4.8E-05        &104021 &6.9E-05        &30550  &2.0E-05        &126791 &8.4E-05\\
0.0016	&49989  &3.3E-05        &91747  &6.1E-05        &36228  &2.4E-05        &115196 &7.6E-05\\
0.002	&32747  &2.2E-05        &74888  &5.0E-05        &33812  &2.2E-05        &97099  &6.4E-05\\
0.004	&3766   &2.5E-06        &8356   &5.5E-06        &30511  &2.0E-05        &21602  &1.4E-05\\
0.006	&347    &2.3E-07        &1397   &9.2E-07        &43555  &2.9E-05        &4867   &3.2E-06\\
0.008	&203    &1.3E-07        &730    &4.8E-07        &3901   &2.6E-06        &2379   &1.6E-06\\
0.012	&107    &7.1E-08        &365    &2.4E-07        &1729   &1.1E-06        &438    &2.9E-07\\
0.016	&60     &4.0E-08        &276    &1.8E-07        &1260   &8.3E-07        &311    &2.1E-07\\
0.02	&38     &2.5E-08        &157    &1.0E-07        &867    &5.7E-07        &117    &7.7E-08\\

\bottomrule
		\end{tabular}
	\begin{flushleft}
{\small Table \ref{tab:mergersC}. Column 1: metallicity; column 2-3: total number of merging BHBs and number of merging BHBs per unit mass at each metallicity for simulations \mobse{}1\_D1.5; column 4-5: same for \mobse{}1\_D; column 6-7: same for \mobse{}1\_D0.02;column 8-9: same for \mobse{}1\_R.}
	\end{flushleft}
	\end{center}
\end{table*}
\begin{table*}
	\begin{center}
		\caption{{ Merging systems in \mobse{}2.}\label{tab:mergersB}}
		\begin{tabular}[t]{lcccccccc}
			\toprule
			{ Z} & \multicolumn{2}{c}{ \mobse{}2\_D1.5}  & \multicolumn{2}{c}{ \mobse{}2\_D} & \multicolumn{2}{c}{ \mobse{}2\_D0.02} & \multicolumn{2}{c}{ \mobse{}2\_R}\\
			 \cmidrule(lr){2-3}\cmidrule(lr){4-5}\cmidrule(lr){6-7}\cmidrule(lr){8-9}
			& { N$_{\mathrm{merger}}$} & { R$_{\rm{cor}}$ [\msun$^{-1}$]} & { N$_{\mathrm{merger}}$} & { R$_{\rm{cor}}$ [\msun$^{-1}$]} & { N$_{\mathrm{merger}}$} & { R$_{\rm{cor}}$ [\msun$^{-1}$]} & { N$_{\mathrm{merger}}$} & { R$_{\rm{cor}}$ [\msun$^{-1}$]}\\
			\midrule
0.0002	&142912 &9.5E-05        &89309  &5.9E-05        &19491  &1.3E-05        &100555 &6.7E-05\\
0.0004	&124392 &8.2E-05        &103508 &6.9E-05        &29680  &2.0E-05        &117595 &7.8E-05\\
0.0008	&92003  &6.1E-05        &97539  &6.5E-05        &30338  &2.0E-05        &118335 &7.8E-05\\
0.0012	&66391  &4.4E-05        &91225  &6.0E-05        &32842  &2.2E-05        &113580 &7.5E-05\\
0.0016	&46703  &3.1E-05        &81349  &5.4E-05        &37037  &2.5E-05        &104239 &6.9E-05\\
0.002	&31499  &2.1E-05        &68936  &4.6E-05        &34330  &2.3E-05        &90655  &6.0E-05\\
0.004	&4030   &2.7E-06        &9472   &6.3E-06        &27772  &1.8E-05        &22906  &1.5E-05\\
0.006	&438    &2.9E-07        &1351   &8.9E-07        &43231  &2.9E-05        &4821   &3.2E-06\\
0.008	&197    &1.3E-07        &722    &4.8E-07        &3598   &2.4E-06        &2414   &1.6E-06\\
0.012	&122    &8.1E-08        &369    &2.4E-07        &1567   &1.0E-06        &406    &2.7E-07\\
0.016	&60     &4.0E-08        &240    &1.6E-07        &1121   &7.4E-07        &259    &1.7E-07\\
0.02	&34     &2.3E-08        &153    &1.0E-07        &869    &5.8E-07        &138    &9.1E-08\\

			\bottomrule
		\end{tabular}
	\begin{flushleft}
{\small Same as Table \ref{tab:mergersC} but for simulations \mobse{}2\_D1.5, \mobse{}2\_D, \mobse{}2\_D0.02 and \mobse{}2\_R.}
	\end{flushleft}
	\end{center}
\end{table*}

\subsection{Formation channels of merging BHBs}
\label{sec:4.4}
In Tables \ref{tab:channels_C} and \ref{tab:channels_B}  we report the most common evolutionary pathways followed by merging BHBs in simulations \mobse{}1 and \mobse{}2, respectively. We distinguish three different formation channels: systems that pass through a single CE phase (SCE); systems which experience multi-CE phases (MCE); and systems that merge without CE phase (ZCE).

The vast majority of progenitors of merging BHBs undergo one CE phase ($>80$\%), even if, especially at low metallicity ($Z<0.004$), about $10-20$ \% of progenitors of merging BHBs do not experience any CE. Merging BHBs that went through multiple CE phases are a negligible fraction.

The percentage of ZCE systems increases with decreasing metallicity in both \mobse{}1 and \mobse{}2. The same trend was also noted by \citet{Dominik2012}, but our results show a stronger dependence on the metallicity than \citet{Dominik2012}.

\section{Conclusions}
\label{sec:5}

We present the \mobse{}~code, our upgraded version of \bse{}. \mobse{} includes up-to-date prescriptions for core-collapse SNe and for stellar winds. We account not only for the metallicity dependence of mass loss, but also for the effect of the Eddington factor \citep{Vink2011,Chen2015,Vink2016}. We discuss two versions of \mobse{}, \mobse{}1 and \mobse{}2. \mobse{}2 implements the metallicity-dependent prescriptions described in \cite{Belczynski2010} and does not account for the effect of the Eddington factor, while \mobse{}1 updates these prescriptions by accounting also for the Eddington factor (following \citealt{Chen2015}). Both versions of \mobse{} also include recipes for PISNe and PPISNe, following \cite{Spera2017}.

The most massive BHs in \mobse{}1 form at low metallicity ($Z\sim{}0.0002-0.002$), reach a mass of $\sim{}50-65$ M$_\odot$ and come from progenitor stars with ZAMS mass $M_{\rm ZAMS}\sim{}60-80$ M$_\odot$. In contrast, the most massive BHs in  \mobse{}2 form  at low metallicity ($Z\sim{}0.0002-0.002$), but reach a lower mass of $\sim{}40$ M$_\odot$ and come from progenitor stars with ZAMS mass $M_{\rm ZAMS}\gtrsim{}120$ M$_\odot$ (Figure~\ref{fig:massspec}).

The distribution of BH masses derived with \mobse2{} is in good agreement with the one discussed by \cite{Belczynski2010,Belczynski2016}, who adopt similar prescriptions for stellar winds and SNe. The distribution of BH masses obtained with \mobse{}1 is significantly different from that produced by \mobse{}2 and is remarkably similar to the one derived with the \sevn{} code \citep{Spera2017}. This is not surprising, because both \mobse{}1 and \sevn{} account for the impact of the Eddington factor on mass loss, but is quite remarkable, because stellar evolution in \mobse{} is still based on the polynomial fitting formulas described in \cite{Hurley2000}, while \sevn{} adopts the recent {\sc PARSEC} stellar evolution models \citep{Bressan2012, Chen2015}. This result also indicates that the Eddington factor has a large impact on the distribution of BH masses.

We have studied the demography of BHBs, performing a large set of population-synthesis simulations  with both \mobse{}1 and \mobse{}2. We perform simulations with different models of core-collapse SN (delayed versus rapid models, \citealt{Fryer2012}) and changing the efficiency of CE.

The distribution of simulated BHB masses covers the entire mass spectrum of BHs predicted by \mobse{}1 and \mobse{}2 from single stellar evolution (Fig.~\ref{fig:massspec}). However, if we consider only merging BHBs (defined as BHBs which are expected to merge in a Hubble time), their maximum mass is significantly lower. Even at the lowest metallicity, the maximum mass of merging BHBs is $\sim{}55$ M$_\odot$ and $\sim{}45$ M$_\odot$ in \mobse{}1 and in \mobse{}2, respectively. This indicates that the most massive BHBs in \mobse{}1 do not merge. The most likely explanation is that these BHBs come from massive ($\sim{}60-80$ M$_\odot$) progenitors which die as red super-giant stars. Thus, all such BHBs form with very large semi-major axis (otherwise their progenitors merge prematurely, due to their large radii) and do not merge in a Hubble time. This feature is nearly independent of the CE parameters and of the SN model.

The maximum mass of merging BHs formed in our simulations ($\approx{}45$ M$_\odot$) is consistent with the possible upper mass gap suggested by LIGO-Virgo detections (i.e. the dearth of merging BHs with mass in the $\sim{}50-135$ M$_\odot$ range, \citealt{Fishbach2017}).

We find merging BHBs with mass ratios in the $0.1 - 1.0$ range, even if mass ratios $>0.6$ are more likely. The masses of our merging BHBs match those of the four observed GW  events. In our \mobse{}1 and \mobse{}2 simulations, systems like GW150914, LVT151012, GW151226, GW170104 and GW170814 could have formed only from binaries with metallicity $Z\leq{}0.006$, $\leq{}0.016$, $\leq{}0.02$, $\leq{}0.012$ and $\leq{}0.008$, respectively. 

The vast majority of progenitors of merging BHBs undergo one CE phase ($>80$\%), even if, especially at low metallicity ($Z<0.004$), about $10-20$ \% of progenitors of merging BHBs do not experience any CE. Merging BHBs that went through multiple CE phases are a negligible fraction.

Merging BHBs form much more efficiently from metal-poor  than from metal-rich binaries, both in \mobse{}1 and in \mobse{}2. The number of BHB mergers per unit mass is $\sim{}10^{-4}$  M$_\odot^{-1}$ at low metallicity ($Z=0.0002-0.002$) and drops to $\sim{}10^{-7}$  M$_\odot^{-1}$ at high metallicity ($Z\sim{}0.02$, Fig.~\ref{fig:rate}). This trend of the number of BHB mergers per unit mass with the progenitor's metallicity potentially has a crucial impact on GW observations across cosmic time.


\begin{table}
	\begin{center}
		\caption{{ Formation channels of BHBs in \mobse{}1\_D.}\label{tab:channels_C}}
		\begin{tabular}[c]{cccc}
			\toprule 
			{ Z} & { N$^{\circ}$ of mergers} & { Channels} & { Fraction} \\ \midrule
			& & ZCE & $0\%$ \\
			0.02 & $157$ & SCE & $100\%$ \\
			& & MCE & $0\%$ \\
			\cmidrule(lr){1-4}
			& & ZCE & $0.36\%$ \\
			0.016 & $276$ & SCE & $98.19\%$ \\
			& & MCE & $1.45\%$ \\
			\cmidrule(lr){1-4}
			& & ZCE & $0.27\%$ \\
			0.012 & $365$ & SCE & $95.89\%$ \\
			& & MCE & $3.84\%$ \\
			\cmidrule(lr){1-4}
			& & ZCE & $0.41\%$ \\
			0.008 & $730$ & SCE & $95.21\%$ \\
			& & MCE & $4.38\%$ \\
			\cmidrule(lr){1-4}
			& & ZCE & $1.00\%$ \\
			0.006 & $1397$ & SCE & $95.71\%$ \\
			& & MCE & $3.29\%$ \\
			\cmidrule(lr){1-4}
			& & ZCE & $17.52\%$ \\
			0.004 & $8356$ & SCE & $81.08\%$ \\
			& & MCE & $1.40\%$ \\
			\cmidrule(lr){1-4}
			& & ZCE & $8.42\%$ \\
			0.002 & $74888$ & SCE & $91.01\%$ \\
			& & MCE & $0.57\%$ \\
			\cmidrule(lr){1-4}
			& & ZCE & $7.94\%$ \\
			0.0016 & $91747$ & SCE & $91.34\%$ \\
			& & MCE & $0.72\%$ \\
			\cmidrule(lr){1-4}
			& & ZCE & $8.13\%$ \\
			0.0012 & $104021$ & SCE & $90.73\%$ \\
			& & MCE & $1.14\%$ \\
			\cmidrule(lr){1-4}
			& & ZCE & $8.97\%$ \\
			0.0008 & $119200$ & SCE & $89.43\%$ \\
			& & MCE & $1.60\%$ \\
			\cmidrule(lr){1-4}
			& & ZCE & $9.16\%$ \\
			0.0004 & $152054$ & SCE & $89.99\%$ \\
			& & MCE & $0.86\%$ \\
			\cmidrule(lr){1-4}
			& & ZCE & $12.68\%$ \\
			0.0002 & $131819$ & SCE & $86.35\%$ \\
			& & MCE & $0.97\%$ \\
			\bottomrule 
		\end{tabular}
\end{center}
	{\small Column 1: metallicity; column 2: total number of BHBs that merge within an Hubble time; column 3: formation channels, considering systems that evolve with zero CE phase (ZCE), with single CE phase (SCE) and with multiple CE phases (MCE); column 4: percentage of the merging BHBs which evolve through a given channel.}
\end{table}

\begin{table}
\begin{center}
		\caption{{ Formation channels of BHBs in \mobse{}2\_D.}\label{tab:channels_B}}
		\begin{tabular}[c]{cccc}
			\toprule
			{ Z} & { N$^{\circ}$ of mergers} & { Channels} & { Fraction} \\ \midrule
			& & ZCE & $0.65\%$ \\
			0.02 & $153$ & SCE & $96.73\%$ \\
			& & MCE & $2.61\%$ \\
			\cmidrule(lr){1-4}
			& & ZCE & $0.00\%$ \\
			0.016 & $240$ & SCE & $96.67\%$ \\
			& & MCE & $3.33\%$ \\
			\cmidrule(lr){1-4}
			& & ZCE & $0.27\%$ \\
			0.012 & $369$ & SCE & $96.75\%$ \\
			& & MCE & $2.98\%$ \\
			\cmidrule(lr){1-4}
			& & ZCE & $0.83\%$ \\
			0.008 & $722$ & SCE & $94.32\%$ \\
			& & MCE & $4.85\%$ \\
			\cmidrule(lr){1-4}
			& & ZCE & $2.74\%$ \\
			0.006 & $1351$ & SCE & $92.23\%$ \\
			& & MCE & $5.03\%$ \\
			\cmidrule(lr){1-4}
			& & ZCE & $16.69\%$ \\
			0.004 & $9472$ & SCE & $82.19\%$ \\
			& & MCE & $1.12\%$ \\
			\cmidrule(lr){1-4}
			& & ZCE & $9.66\%$ \\
			0.002 & $68936$ & SCE & $89.91\%$ \\
			& & MCE & $0.43\%$ \\
			\cmidrule(lr){1-4}
			& & ZCE & $9.62\%$ \\
			0.0016 & $81349$ & SCE & $89.90\%$ \\
			& & MCE & $0.48\%$ \\
			\cmidrule(lr){1-4}
			& & ZCE & $10.14\%$ \\
			0.0012 & $91225$ & SCE & $88.61\%$ \\
			& & MCE & $1.25\%$ \\
			\cmidrule(lr){1-4}
			& & ZCE & $11.71\%$ \\
			0.0008 & $9705$ & SCE & $86.37\%$ \\
			& & MCE & $1.93\%$ \\
			\cmidrule(lr){1-4}
			& & ZCE & $13.71\%$ \\
			0.0004 & $103508$ & SCE & $85.07\%$ \\
			& & MCE & $1.22\%$ \\
			\cmidrule(lr){1-4}
			& & ZCE & $17.46\%$ \\
			0.0002 & $89309$ & SCE & $81.13\%$ \\
			& & MCE & $1.41\%$ \\
			\bottomrule
		\end{tabular}
	\end{center}
	{\small Same as Table~\ref{tab:channels_C} but for \mobse2.}
\end{table}


\section*{Acknowledgements}
We thank the anonymous referee for their comments which significantly improved this paper.  We thank Alessandro Bressan for useful discussions.
We acknowledge the "Accordo Quadro INAF-CINECA (2017)" for the availability of high performance computing resources and support. 
MM and MS acknowledge financial support from the Italian Ministry of Education, University and Research (MIUR) through grant FIRB 2012 RBFR12PM1F, and from INAF through grant PRIN-2014-14. MM acknowledges financial support from the MERAC Foundation.




\bibliographystyle{mnras}
\bibliography{biblio} 

\begin{thebibliography}{}
\makeatletter
\relax
\def\mn@urlcharsother{\let\do\@makeother \do\$\do\&\do\#\do\^\do\_\do\%\do\~}
\def\mn@doi{\begingroup\mn@urlcharsother \@ifnextchar [ {\mn@doi@}
  {\mn@doi@[]}}
\def\mn@doi@[#1]#2{\def\@tempa{#1}\ifx\@tempa\@empty \href
  {http://dx.doi.org/#2} {doi:#2}\else \href {http://dx.doi.org/#2} {#1}\fi
  \endgroup}
\def\mn@eprint#1#2{\mn@eprint@#1:#2::\@nil}
\def\mn@eprint@arXiv#1{\href {http://arxiv.org/abs/#1} {{\tt arXiv:#1}}}
\def\mn@eprint@dblp#1{\href {http://dblp.uni-trier.de/rec/bibtex/#1.xml}
  {dblp:#1}}
\def\mn@eprint@#1:#2:#3:#4\@nil{\def\@tempa {#1}\def\@tempb {#2}\def\@tempc
  {#3}\ifx \@tempc \@empty \let \@tempc \@tempb \let \@tempb \@tempa \fi \ifx
  \@tempb \@empty \def\@tempb {arXiv}\fi \@ifundefined
  {mn@eprint@\@tempb}{\@tempb:\@tempc}{\expandafter \expandafter \csname
  mn@eprint@\@tempb\endcsname \expandafter{\@tempc}}}

\bibitem[\protect\citeauthoryear{{Abbott} et~al.,}{{Abbott}
  et~al.}{2016a}]{Abbott2016a}
{Abbott} B.~P.,  et~al., 2016a, \mn@doi [Physical Review X]
  {10.1103/PhysRevX.6.041015}, \href
  {http://adsabs.harvard.edu/abs/2016PhRvX...6d1015A} {6, 041015}

\bibitem[\protect\citeauthoryear{{Abbott} et~al.,}{{Abbott}
  et~al.}{2016b}]{Abbott2016b}
{Abbott} B.~P.,  et~al., 2016b, \mn@doi [Physical Review Letters]
  {10.1103/PhysRevLett.116.061102}, \href
  {http://adsabs.harvard.edu/abs/2016PhRvL.116f1102A} {116, 061102}

\bibitem[\protect\citeauthoryear{{Abbott} et~al.,}{{Abbott}
  et~al.}{2016c}]{Abbott2016c}
{Abbott} B.~P.,  et~al., 2016c, \mn@doi [\apjl] {10.3847/2041-8205/818/2/L22},
  \href {http://adsabs.harvard.edu/abs/2016ApJ...818L..22A} {818, L22}

\bibitem[\protect\citeauthoryear{{Abbott} et~al.,}{{Abbott}
  et~al.}{2017}]{Abbott2017}
{Abbott} B.~P.,  et~al., 2017, \mn@doi [Physical Review Letters]
  {10.1103/PhysRevLett.118.221101}, \href
  {http://adsabs.harvard.edu/abs/2017PhRvL.118v1101A} {118, 221101}

\bibitem[\protect\citeauthoryear{{Antonini}, {Toonen}  \& {Hamers}}{{Antonini}
  et~al.}{2017}]{Antonini2017}
{Antonini} F.,  {Toonen} S.,   {Hamers} A.~S.,  2017, \mn@doi [\apj]
  {10.3847/1538-4357/aa6f5e}, \href
  {http://adsabs.harvard.edu/abs/2017ApJ...841...77A} {841, 77}

\bibitem[\protect\citeauthoryear{{Askar}, {Giersz}, {Pych}, {Olech}  \&
  {Hypki}}{{Askar} et~al.}{2016}]{Askar2016}
{Askar} A.,  {Giersz} M.,  {Pych} W.,  {Olech} A.,   {Hypki} A.,  2016, in
  {Meiron} Y.,  {Li} S.,  {Liu} F.-K.,   {Spurzem} R.,  eds,  IAU Symposium
  Vol. 312, Star Clusters and Black Holes in Galaxies across Cosmic Time. pp
  262--263 (\mn@eprint {arXiv} {1501.00417}),
  \mn@doi{10.1017/S1743921315007991}

\bibitem[\protect\citeauthoryear{{Banerjee}}{{Banerjee}}{2017}]{Banerjee2017}
{Banerjee} S.,  2017, preprint, \href
  {http://adsabs.harvard.edu/abs/2017arXiv170700922B} {} (\mn@eprint {arXiv}
  {1707.00922})

\bibitem[\protect\citeauthoryear{{Barkat}, {Rakavy}  \& {Sack}}{{Barkat}
  et~al.}{1967}]{Barkat1967}
{Barkat} Z.,  {Rakavy} G.,   {Sack} N.,  1967, \mn@doi [Physical Review
  Letters] {10.1103/PhysRevLett.18.379}, \href
  {http://adsabs.harvard.edu/abs/1967PhRvL..18..379B} {18, 379}

\bibitem[\protect\citeauthoryear{{Belczynski}, {Sadowski}  \&
  {Rasio}}{{Belczynski} et~al.}{2004}]{Belczynski2004}
{Belczynski} K.,  {Sadowski} A.,   {Rasio} F.~A.,  2004, \mn@doi [\apj]
  {10.1086/422191}, \href {http://adsabs.harvard.edu/abs/2004ApJ...611.1068B}
  {611, 1068}

\bibitem[\protect\citeauthoryear{{Belczynski}, {Bulik}, {Fryer}, {Ruiter},
  {Valsecchi}, {Vink}  \& {Hurley}}{{Belczynski} et~al.}{2010}]{Belczynski2010}
{Belczynski} K.,  {Bulik} T.,  {Fryer} C.~L.,  {Ruiter} A.,  {Valsecchi} F.,
  {Vink} J.~S.,   {Hurley} J.~R.,  2010, \apj, 714, 1217

\bibitem[\protect\citeauthoryear{{Belczynski}, {Holz}, {Bulik}  \&
  {O'Shaughnessy}}{{Belczynski} et~al.}{2016a}]{Belczynski2016}
{Belczynski} K.,  {Holz} D.~E.,  {Bulik} T.,   {O'Shaughnessy} R.,  2016a,
  \mn@doi [\nat] {10.1038/nature18322}, \href
  {http://adsabs.harvard.edu/abs/2016Natur.534..512B} {534, 512}

\bibitem[\protect\citeauthoryear{{Belczynski} et~al.,}{{Belczynski}
  et~al.}{2016b}]{Belczynski2016pair}
{Belczynski} K.,  et~al., 2016b, \mn@doi [\aap] {10.1051/0004-6361/201628980},
  \href {http://adsabs.harvard.edu/abs/2016A%26A...594A..97B} {594, A97}

\bibitem[\protect\citeauthoryear{{Bestenlehner} et~al.,}{{Bestenlehner}
  et~al.}{2014}]{Bestenlehner2014}
{Bestenlehner} J.~M.,  et~al., 2014, \mn@doi [\aap]
  {10.1051/0004-6361/201423643}, \href
  {http://adsabs.harvard.edu/abs/2014A%26A...570A..38B} {570, A38}

\bibitem[\protect\citeauthoryear{{Bethe}}{{Bethe}}{1990}]{Bethe1990}
{Bethe} H.~A.,  1990, \mn@doi [Reviews of Modern Physics]
  {10.1103/RevModPhys.62.801}, \href
  {http://adsabs.harvard.edu/abs/1990RvMP...62..801B} {62, 801}

\bibitem[\protect\citeauthoryear{{Bond}, {Arnett}  \& {Carr}}{{Bond}
  et~al.}{1984}]{Bond1984}
{Bond} J.~R.,  {Arnett} W.~D.,   {Carr} B.~J.,  1984, \mn@doi [\apj]
  {10.1086/162057}, \href {http://adsabs.harvard.edu/abs/1984ApJ...280..825B}
  {280, 825}

\bibitem[\protect\citeauthoryear{{Bresolin} \& {Kudritzki}}{{Bresolin} \&
  {Kudritzki}}{2004}]{Bresolin2004}
{Bresolin} F.,  {Kudritzki} R.~P.,  2004, Origin and Evolution of the Elements,
  \href {http://adsabs.harvard.edu/abs/2004oee..symp..283B} {p.~283}

\bibitem[\protect\citeauthoryear{{Bressan}, {Marigo}, {Girardi}, {Salasnich},
  {Dal Cero}, {Rubele}  \& {Nanni}}{{Bressan} et~al.}{2012}]{Bressan2012}
{Bressan} A.,  {Marigo} P.,  {Girardi} L.,  {Salasnich} B.,  {Dal Cero} C.,
  {Rubele} S.,   {Nanni} A.,  2012, \mn@doi [\mnras]
  {10.1111/j.1365-2966.2012.21948.x}, \href
  {http://adsabs.harvard.edu/abs/2012MNRAS.427..127B} {427, 127}

\bibitem[\protect\citeauthoryear{{Burrows}}{{Burrows}}{2013}]{Burrows2013}
{Burrows} A.,  2013, \mn@doi [Reviews of Modern Physics]
  {10.1103/RevModPhys.85.245}, \href
  {http://adsabs.harvard.edu/abs/2013RvMP...85..245B} {85, 245}

\bibitem[\protect\citeauthoryear{{Chen}, {Woosley}, {Heger}, {Almgren}  \&
  {Whalen}}{{Chen} et~al.}{2014}]{Chen2014}
{Chen} K.-J.,  {Woosley} S.,  {Heger} A.,  {Almgren} A.,   {Whalen} D.~J.,
  2014, \mn@doi [\apj] {10.1088/0004-637X/792/1/28}, \href
  {http://adsabs.harvard.edu/abs/2014ApJ...792...28C} {792, 28}

\bibitem[\protect\citeauthoryear{{Chen}, {Bressan}, {Girardi}, {Marigo}, {Kong}
   \& {Lanza}}{{Chen} et~al.}{2015}]{Chen2015}
{Chen} Y.,  {Bressan} A.,  {Girardi} L.,  {Marigo} P.,  {Kong} X.,   {Lanza}
  A.,  2015, \mn@doi [\mnras] {10.1093/mnras/stv1281}, \href
  {http://adsabs.harvard.edu/abs/2015MNRAS.452.1068C} {452, 1068}

\bibitem[\protect\citeauthoryear{{Colpi}, {Mapelli}  \& {Possenti}}{{Colpi}
  et~al.}{2003}]{Colpi2003}
{Colpi} M.,  {Mapelli} M.,   {Possenti} A.,  2003, \mn@doi [\apj]
  {10.1086/379543}, \href {http://adsabs.harvard.edu/abs/2003ApJ...599.1260C}
  {599, 1260}

\bibitem[\protect\citeauthoryear{{Dominik}, {Belczynski}, {Fryer}, {Holz},
  {Berti}, {Bulik}, {Mandel}  \& {O'Shaughnessy}}{{Dominik}
  et~al.}{2012}]{Dominik2012}
{Dominik} M.,  {Belczynski} K.,  {Fryer} C.,  {Holz} D.~E.,  {Berti} E.,
  {Bulik} T.,  {Mandel} I.,   {O'Shaughnessy} R.,  2012, \mn@doi [\apj]
  {10.1088/0004-637X/759/1/52}, \href
  {http://adsabs.harvard.edu/abs/2012ApJ...759...52D} {759, 52}

\bibitem[\protect\citeauthoryear{{Ertl}, {Janka}, {Woosley}, {Sukhbold}  \&
  {Ugliano}}{{Ertl} et~al.}{2016}]{Ertl2016}
{Ertl} T.,  {Janka} H.-T.,  {Woosley} S.~E.,  {Sukhbold} T.,   {Ugliano} M.,
  2016, \mn@doi [\apj] {10.3847/0004-637X/818/2/124}, \href
  {http://adsabs.harvard.edu/abs/2016ApJ...818..124E} {818, 124}

\bibitem[\protect\citeauthoryear{{Farr}, {Sravan}, {Cantrell}, {Kreidberg},
  {Bailyn}, {Mandel}  \& {Kalogera}}{{Farr} et~al.}{2011}]{Farr2011}
{Farr} W.~M.,  {Sravan} N.,  {Cantrell} A.,  {Kreidberg} L.,  {Bailyn} C.~D.,
  {Mandel} I.,   {Kalogera} V.,  2011, \mn@doi [\apj]
  {10.1088/0004-637X/741/2/103}, \href
  {http://adsabs.harvard.edu/abs/2011ApJ...741..103F} {741, 103}

\bibitem[\protect\citeauthoryear{{Fishbach} \& {Holz}}{{Fishbach} \&
  {Holz}}{2017}]{Fishbach2017}
{Fishbach} M.,  {Holz} D.~E.,  2017, preprint, \href
  {http://adsabs.harvard.edu/abs/2017arXiv170908584F} {} (\mn@eprint {arXiv}
  {1709.08584})

\bibitem[\protect\citeauthoryear{{Fryer}}{{Fryer}}{1999}]{Fryer1999}
{Fryer} C.~L.,  1999, \mn@doi [\apj] {10.1086/307647}, \href
  {http://adsabs.harvard.edu/abs/1999ApJ...522..413F} {522, 413}

\bibitem[\protect\citeauthoryear{{Fryer}}{{Fryer}}{2006}]{Fryer2006}
{Fryer} C.~L.,  2006, \mn@doi [\nar] {10.1016/j.newar.2006.06.052}, \href
  {http://adsabs.harvard.edu/abs/2006NewAR..50..492F} {50, 492}

\bibitem[\protect\citeauthoryear{{Fryer}, {Woosley}  \& {Heger}}{{Fryer}
  et~al.}{2001}]{Fryer2001}
{Fryer} C.~L.,  {Woosley} S.~E.,   {Heger} A.,  2001, \mn@doi [\apj]
  {10.1086/319719}, \href {http://adsabs.harvard.edu/abs/2001ApJ...550..372F}
  {550, 372}

\bibitem[\protect\citeauthoryear{{Fryer}, {Belczynski}, {Wiktorowicz},
  {Dominik}, {Kalogera}  \& {Holz}}{{Fryer} et~al.}{2012}]{Fryer2012}
{Fryer} C.~L.,  {Belczynski} K.,  {Wiktorowicz} G.,  {Dominik} M.,  {Kalogera}
  V.,   {Holz} D.~E.,  2012, \mn@doi [\apj] {10.1088/0004-637X/749/1/91}, \href
  {http://adsabs.harvard.edu/abs/2012ApJ...749...91F} {749, 91}

\bibitem[\protect\citeauthoryear{{Gr{\"a}fener} \& {Hamann}}{{Gr{\"a}fener} \&
  {Hamann}}{2008}]{Graefener2008}
{Gr{\"a}fener} G.,  {Hamann} W.-R.,  2008, \mn@doi [\aap]
  {10.1051/0004-6361:20066176}, \href
  {http://adsabs.harvard.edu/abs/2008A%26A...482..945G} {482, 945}

\bibitem[\protect\citeauthoryear{{Gr{\"a}fener}, {Vink}, {de Koter}  \&
  {Langer}}{{Gr{\"a}fener} et~al.}{2011}]{Graefener2011}
{Gr{\"a}fener} G.,  {Vink} J.~S.,  {de Koter} A.,   {Langer} N.,  2011, \mn@doi
  [\aap] {10.1051/0004-6361/201116701}, \href
  {http://adsabs.harvard.edu/abs/2011A%26A...535A..56G} {535, A56}

\bibitem[\protect\citeauthoryear{{Gr{\"a}fener}, {Owocki}  \&
  {Vink}}{{Gr{\"a}fener} et~al.}{2012}]{Graefener2012}
{Gr{\"a}fener} G.,  {Owocki} S.~P.,   {Vink} J.~S.,  2012, \mn@doi [\aap]
  {10.1051/0004-6361/201117497}, \href
  {http://adsabs.harvard.edu/abs/2012A%26A...538A..40G} {538, A40}

\bibitem[\protect\citeauthoryear{{Hall} \& {Tout}}{{Hall} \&
  {Tout}}{2014}]{Hall2014}
{Hall} P.~D.,  {Tout} C.~A.,  2014, \mn@doi [\mnras] {10.1093/mnras/stu1678},
  \href {http://adsabs.harvard.edu/abs/2014MNRAS.444.3209H} {444, 3209}

\bibitem[\protect\citeauthoryear{{Heger} \& {Woosley}}{{Heger} \&
  {Woosley}}{2002}]{Heger2002}
{Heger} A.,  {Woosley} S.~E.,  2002, \mn@doi [\apj] {10.1086/338487}, \href
  {http://adsabs.harvard.edu/abs/2002ApJ...567..532H} {567, 532}

\bibitem[\protect\citeauthoryear{{Heger}, {Fryer}, {Woosley}, {Langer}  \&
  {Hartmann}}{{Heger} et~al.}{2003}]{Heger2003}
{Heger} A.,  {Fryer} C.~L.,  {Woosley} S.~E.,  {Langer} N.,   {Hartmann} D.~H.,
   2003, \mn@doi [\apj] {10.1086/375341}, \href
  {http://adsabs.harvard.edu/abs/2003ApJ...591..288H} {591, 288}

\bibitem[\protect\citeauthoryear{{Hobbs}, {Lorimer}, {Lyne}  \&
  {Kramer}}{{Hobbs} et~al.}{2005}]{Hobbs2005}
{Hobbs} G.,  {Lorimer} D.~R.,  {Lyne} A.~G.,   {Kramer} M.,  2005, \mn@doi
  [\mnras] {10.1111/j.1365-2966.2005.09087.x}, \href
  {http://adsabs.harvard.edu/abs/2005MNRAS.360..974H} {360, 974}

\bibitem[\protect\citeauthoryear{{Hurley}, {Pols}  \& {Tout}}{{Hurley}
  et~al.}{2000}]{Hurley2000}
{Hurley} J.~R.,  {Pols} O.~R.,   {Tout} C.~A.,  2000, \mn@doi [MNRAS]
  {10.1046/j.1365-8711.2000.03426.x}, \href
  {http://adsabs.harvard.edu/abs/2000MNRAS.315..543H} {315, 543}

\bibitem[\protect\citeauthoryear{{Hurley}, {Tout}  \& {Pols}}{{Hurley}
  et~al.}{2002}]{Hurley2002}
{Hurley} J.~R.,  {Tout} C.~A.,   {Pols} O.~R.,  2002, \mn@doi [MNRAS]
  {10.1046/j.1365-8711.2002.05038.x}, \href
  {http://adsabs.harvard.edu/abs/2002MNRAS.329..897H} {329, 8 97}

\bibitem[\protect\citeauthoryear{{Ivanova} \& {Taam}}{{Ivanova} \&
  {Taam}}{2004}]{Ivanova2004}
{Ivanova} N.,  {Taam} R.~E.,  2004, \mn@doi [\apj] {10.1086/380561}, \href
  {http://adsabs.harvard.edu/abs/2004ApJ...601.1058I} {601, 1058}

\bibitem[\protect\citeauthoryear{{Ivanova} et~al.,}{{Ivanova}
  et~al.}{2013}]{Ivanova2013}
{Ivanova} N.,  et~al., 2013, \mn@doi [\aapr] {10.1007/s00159-013-0059-2}, \href
  {http://adsabs.harvard.edu/abs/2013A%26ARv..21...59I} {21, 59}

\bibitem[\protect\citeauthoryear{{Janka}}{{Janka}}{2012}]{Janka2012}
{Janka} H.-T.,  2012, \mn@doi [Annual Review of Nuclear and Particle Science]
  {10.1146/annurev-nucl-102711-094901}, \href
  {http://adsabs.harvard.edu/abs/2012ARNPS..62..407J} {62, 407}

\bibitem[\protect\citeauthoryear{{Janka}, {Langanke}, {Marek},
  {Mart{\'{\i}}nez-Pinedo}  \& {M{\"u}ller}}{{Janka} et~al.}{2007}]{Janka2007}
{Janka} H.-T.,  {Langanke} K.,  {Marek} A.,  {Mart{\'{\i}}nez-Pinedo} G.,
  {M{\"u}ller} B.,  2007, \mn@doi [\physrep] {10.1016/j.physrep.2007.02.002},
  \href {http://adsabs.harvard.edu/abs/2007PhR...442...38J} {442, 38}

\bibitem[\protect\citeauthoryear{{Kroupa}}{{Kroupa}}{2001}]{Kroupa2001}
{Kroupa} P.,  2001, \mn@doi [\mnras] {10.1046/j.1365-8711.2001.04022.x}, \href
  {http://adsabs.harvard.edu/abs/2001MNRAS.322..231K} {322, 231}

\bibitem[\protect\citeauthoryear{{Kudritzki}}{{Kudritzki}}{2002}]{Kudritzki2002}
{Kudritzki} R.~P.,  2002, \mn@doi [\apj] {10.1086/342178}, \href
  {http://adsabs.harvard.edu/abs/2002ApJ...577..389K} {577, 389}

\bibitem[\protect\citeauthoryear{{Kulkarni}, {Hut}  \& {McMillan}}{{Kulkarni}
  et~al.}{1993}]{Kulkarni1993}
{Kulkarni} S.~R.,  {Hut} P.,   {McMillan} S.,  1993, \mn@doi [\nat]
  {10.1038/364421a0}, \href {http://adsabs.harvard.edu/abs/1993Natur.364..421K}
  {364, 421}

\bibitem[\protect\citeauthoryear{{Limongi}}{{Limongi}}{2017}]{Limongi2017}
{Limongi} M.,  2017, preprint, \href
  {http://adsabs.harvard.edu/abs/2017arXiv170601913L} {} (\mn@eprint {arXiv}
  {1706.01913})

\bibitem[\protect\citeauthoryear{{Loveridge}, {van der Sluys}  \&
  {Kalogera}}{{Loveridge} et~al.}{2011}]{Loveridge2011}
{Loveridge} A.~J.,  {van der Sluys} M.~V.,   {Kalogera} V.,  2011, \mn@doi
  [\apj] {10.1088/0004-637X/743/1/49}, \href
  {http://adsabs.harvard.edu/abs/2011ApJ...743...49L} {743, 49}

\bibitem[\protect\citeauthoryear{{Maggiore}}{{Maggiore}}{2008}]{Maggiore2008}
{Maggiore} M.,  2008, \mn@doi [Classical and Quantum Gravity]
  {10.1088/0264-9381/25/20/209002}, \href
  {http://adsabs.harvard.edu/abs/2008CQGra..25t9002.} {25, 209002}

\bibitem[\protect\citeauthoryear{{Mandel} \& {de Mink}}{{Mandel} \& {de
  Mink}}{2016}]{Mandel2016}
{Mandel} I.,  {de Mink} S.~E.,  2016, \mn@doi [\mnras] {10.1093/mnras/stw379},
  \href {http://adsabs.harvard.edu/abs/2016MNRAS.458.2634M} {458, 2634}

\bibitem[\protect\citeauthoryear{{Mapelli}}{{Mapelli}}{2016}]{Mapelli2016}
{Mapelli} M.,  2016, \mn@doi [\mnras] {10.1093/mnras/stw869}, \href
  {http://adsabs.harvard.edu/abs/2016MNRAS.459.3432M} {459, 3432}

\bibitem[\protect\citeauthoryear{{Mapelli}, {Colpi}  \& {Zampieri}}{{Mapelli}
  et~al.}{2009}]{Mapelli2009}
{Mapelli} M.,  {Colpi} M.,   {Zampieri} L.,  2009, \mn@doi [MNRAS]
  {10.1111/j.1745-3933.2009.00645.x}, \href
  {http://adsabs.harvard.edu/abs/2009MNRAS.395L..71M} {395, L71}

\bibitem[\protect\citeauthoryear{{Mapelli}, {Ripamonti}, {Zampieri}, {Colpi}
  \& {Bressan}}{{Mapelli} et~al.}{2010}]{Mapelli2010}
{Mapelli} M.,  {Ripamonti} E.,  {Zampieri} L.,  {Colpi} M.,   {Bressan} A.,
  2010, \mn@doi [\mnras] {10.1111/j.1365-2966.2010.17048.x}, \href
  {http://adsabs.harvard.edu/abs/2010MNRAS.408..234M} {408, 234}

\bibitem[\protect\citeauthoryear{{Mapelli}, {Zampieri}, {Ripamonti}  \&
  {Bressan}}{{Mapelli} et~al.}{2013}]{Mapelli2013}
{Mapelli} M.,  {Zampieri} L.,  {Ripamonti} E.,   {Bressan} A.,  2013, \mn@doi
  [\mnras] {10.1093/mnras/sts500}, \href
  {http://adsabs.harvard.edu/abs/2013MNRAS.429.2298M} {429, 2298}

\bibitem[\protect\citeauthoryear{{Mapelli}, {Giacobbo}, {Ripamonti}  \&
  {Spera}}{{Mapelli} et~al.}{2017}]{Mapelli2017}
{Mapelli} M.,  {Giacobbo} N.,  {Ripamonti} E.,   {Spera} M.,  2017, preprint,
  \href {http://adsabs.harvard.edu/abs/2017arXiv170805722M} {} (\mn@eprint
  {arXiv} {1708.05722})

\bibitem[\protect\citeauthoryear{{Marchant}, {Langer}, {Podsiadlowski},
  {Tauris}  \& {Moriya}}{{Marchant} et~al.}{2016}]{Marchant2016}
{Marchant} P.,  {Langer} N.,  {Podsiadlowski} P.,  {Tauris} T.~M.,   {Moriya}
  T.~J.,  2016, \mn@doi [\aap] {10.1051/0004-6361/201628133}, \href
  {http://adsabs.harvard.edu/abs/2016A%26A...588A..50M} {588, A50}

\bibitem[\protect\citeauthoryear{{Meynet} \& {Maeder}}{{Meynet} \&
  {Maeder}}{2005}]{Meynet2005}
{Meynet} G.,  {Maeder} A.,  2005, \mn@doi [\aap] {10.1051/0004-6361:20047106},
  \href {http://adsabs.harvard.edu/abs/2005A%26A...429..581M} {429, 581}

\bibitem[\protect\citeauthoryear{{Muijres}, {Vink}, {de Koter}, {M{\"u}ller}
  \& {Langer}}{{Muijres} et~al.}{2012}]{Muijres2012}
{Muijres} L.~E.,  {Vink} J.~S.,  {de Koter} A.,  {M{\"u}ller} P.~E.,   {Langer}
  N.,  2012, \mn@doi [\aap] {10.1051/0004-6361/201015818}, \href
  {http://adsabs.harvard.edu/abs/2012A%26A...537A..37M} {537, A37}

\bibitem[\protect\citeauthoryear{{O'Connor} \& {Ott}}{{O'Connor} \&
  {Ott}}{2011}]{Oconnor2011}
{O'Connor} E.,  {Ott} C.~D.,  2011, \mn@doi [\apj]
  {10.1088/0004-637X/730/2/70}, \href
  {http://adsabs.harvard.edu/abs/2011ApJ...730...70O} {730, 70}

\bibitem[\protect\citeauthoryear{{Ober}, {El Eid}  \& {Fricke}}{{Ober}
  et~al.}{1983}]{Ober1983}
{Ober} W.~W.,  {El Eid} M.~F.,   {Fricke} K.~J.,  1983, \aap, \href
  {http://adsabs.harvard.edu/abs/1983A%26A...119...61O} {119, 61}

\bibitem[\protect\citeauthoryear{{Oppenheimer} \& {Volkoff}}{{Oppenheimer} \&
  {Volkoff}}{1939}]{Oppenheimer1939}
{Oppenheimer} J.~R.,  {Volkoff} G.~M.,  1939, \mn@doi [Physical Review]
  {10.1103/PhysRev.55.374}, \href
  {http://adsabs.harvard.edu/abs/1939PhRv...55..374O} {55, 374}

\bibitem[\protect\citeauthoryear{{{\"O}zel}, {Psaltis}, {Narayan}  \&
  {McClintock}}{{{\"O}zel} et~al.}{2010}]{Ozel2010}
{{\"O}zel} F.,  {Psaltis} D.,  {Narayan} R.,   {McClintock} J.~E.,  2010,
  \mn@doi [\apj] {10.1088/0004-637X/725/2/1918}, \href
  {http://adsabs.harvard.edu/abs/2010ApJ...725.1918O} {725, 1918}

\bibitem[\protect\citeauthoryear{{Pejcha} \& {Prieto}}{{Pejcha} \&
  {Prieto}}{2015}]{Pejcha2015}
{Pejcha} O.,  {Prieto} J.~L.,  2015, \mn@doi [\apj]
  {10.1088/0004-637X/806/2/225}, \href
  {http://adsabs.harvard.edu/abs/2015ApJ...806..225P} {806, 225}

\bibitem[\protect\citeauthoryear{{Petrov}, {Vink}  \& {Gr{\"a}fener}}{{Petrov}
  et~al.}{2016}]{Petrov2016}
{Petrov} B.,  {Vink} J.~S.,   {Gr{\"a}fener} G.,  2016, \mn@doi [\mnras]
  {10.1093/mnras/stw382}, \href
  {http://adsabs.harvard.edu/abs/2016MNRAS.458.1999P} {458, 1999}

\bibitem[\protect\citeauthoryear{{Portegies Zwart} \& {McMillan}}{{Portegies
  Zwart} \& {McMillan}}{2000}]{Portegies2000}
{Portegies Zwart} S.~F.,  {McMillan} S.~L.~W.,  2000, \mn@doi [\apjl]
  {10.1086/312422}, \href {http://adsabs.harvard.edu/abs/2000ApJ...528L..17P}
  {528, L17}

\bibitem[\protect\citeauthoryear{{Rodriguez}, {Morscher}, {Pattabiraman},
  {Chatterjee}, {Haster}  \& {Rasio}}{{Rodriguez} et~al.}{2015}]{Rodriguez2015}
{Rodriguez} C.~L.,  {Morscher} M.,  {Pattabiraman} B.,  {Chatterjee} S.,
  {Haster} C.-J.,   {Rasio} F.~A.,  2015, \mn@doi [Physical Review Letters]
  {10.1103/PhysRevLett.115.051101}, \href
  {http://adsabs.harvard.edu/abs/2015PhRvL.115e1101R} {115, 051101}

\bibitem[\protect\citeauthoryear{{Rodriguez}, {Zevin}, {Pankow}, {Kalogera}  \&
  {Rasio}}{{Rodriguez} et~al.}{2016}]{Rodriguez2016}
{Rodriguez} C.~L.,  {Zevin} M.,  {Pankow} C.,  {Kalogera} V.,   {Rasio} F.~A.,
  2016, \mn@doi [\apjl] {10.3847/2041-8205/832/1/L2}, \href
  {http://adsabs.harvard.edu/abs/2016ApJ...832L...2R} {832, L2}

\bibitem[\protect\citeauthoryear{{Sana} et~al.,}{{Sana}
  et~al.}{2012}]{Sana2012}
{Sana} H.,  et~al., 2012, \mn@doi [Science] {10.1126/science.1223344}, \href
  {http://adsabs.harvard.edu/abs/2012Sci...337..444S} {337, 444}

\bibitem[\protect\citeauthoryear{{Sana} et~al.,}{{Sana}
  et~al.}{2013}]{Sana2013}
{Sana} H.,  et~al., 2013, \mn@doi [\aap] {10.1051/0004-6361/201219621}, \href
  {http://adsabs.harvard.edu/abs/2013A%26A...550A.107S} {550, A107}

\bibitem[\protect\citeauthoryear{{Schutz}}{{Schutz}}{1989}]{Schutz1989}
{Schutz} B.~F.,  1989, in {Hellings} R.~W.,  ed.,  NASA Conference Publication
  Vol. 3046, NASA Conference Publication.

\bibitem[\protect\citeauthoryear{{Sigurdsson} \& {Phinney}}{{Sigurdsson} \&
  {Phinney}}{1993}]{Sigurdsson1993}
{Sigurdsson} S.,  {Phinney} E.~S.,  1993, \mn@doi [\apj] {10.1086/173190},
  \href {http://adsabs.harvard.edu/abs/1993ApJ...415..631S} {415, 631}

\bibitem[\protect\citeauthoryear{{Smartt}}{{Smartt}}{2009}]{Smartt2009}
{Smartt} S.~J.,  2009, \mn@doi [Annual Review of Astronomy \& Astrophysics]
  {10.1146/annurev-astro-082708-101737}, \href
  {http://adsabs.harvard.edu/abs/2009ARA%26A..47...63S} {47, 63}

\bibitem[\protect\citeauthoryear{{Spera} \& {Mapelli}}{{Spera} \&
  {Mapelli}}{2017}]{Spera2017}
{Spera} M.,  {Mapelli} M.,  2017, \mn@doi [\mnras] {10.1093/mnras/stx1576},
  \href {http://adsabs.harvard.edu/abs/2017MNRAS.470.4739S} {470, 4739}

\bibitem[\protect\citeauthoryear{{Spera}, {Mapelli}  \& {Bressan}}{{Spera}
  et~al.}{2015}]{Spera2015}
{Spera} M.,  {Mapelli} M.,   {Bressan} A.,  2015, \mn@doi [MNRAS]
  {10.1093/mnras/stv1161}, \href
  {http://adsabs.harvard.edu/abs/2015MNRAS.451.4086S} {451, 4086}

\bibitem[\protect\citeauthoryear{{Spera}, {Giacobbo}  \& {Mapelli}}{{Spera}
  et~al.}{2016}]{Spera2016}
{Spera} M.,  {Giacobbo} N.,   {Mapelli} M.,  2016, preprint, \href
  {http://adsabs.harvard.edu/abs/2016arXiv160603349S} {} (\mn@eprint {arXiv}
  {1606.03349})

\bibitem[\protect\citeauthoryear{{Tang}, {Bressan}, {Rosenfield}, {Slemer},
  {Marigo}, {Girardi}  \& {Bianchi}}{{Tang} et~al.}{2014}]{Tang2014}
{Tang} J.,  {Bressan} A.,  {Rosenfield} P.,  {Slemer} A.,  {Marigo} P.,
  {Girardi} L.,   {Bianchi} L.,  2014, \mn@doi [MNRAS] {10.1093/mnras/stu2029},
  \href {http://adsabs.harvard.edu/abs/2014MNRAS.445.4287T} {445, 4287}

\bibitem[\protect\citeauthoryear{{The LIGO Scientific Collaboration}
  et~al.,}{{The LIGO Scientific Collaboration} et~al.}{2017}]{LIGO2017}
{The LIGO Scientific Collaboration} et~al., 2017, preprint, \href
  {http://adsabs.harvard.edu/abs/2017arXiv170909660T} {} (\mn@eprint {arXiv}
  {1709.09660})

\bibitem[\protect\citeauthoryear{{Thorne}}{{Thorne}}{1987}]{Thorne1987}
{Thorne} K.~S.,  1987, Science, \href
  {http://adsabs.harvard.edu/abs/1987Sci...236Q1007T} {236, 1007}

\bibitem[\protect\citeauthoryear{{Timmes}, {Woosley}  \& {Weaver}}{{Timmes}
  et~al.}{1996}]{Timmes1996}
{Timmes} F.~X.,  {Woosley} S.~E.,   {Weaver} T.~A.,  1996, \mn@doi [\apj]
  {10.1086/176778}, \href {http://adsabs.harvard.edu/abs/1996ApJ...457..834T}
  {457, 834}

\bibitem[\protect\citeauthoryear{{Toonen}, {Nelemans}  \& {Portegies
  Zwart}}{{Toonen} et~al.}{2012}]{Toonen2012}
{Toonen} S.,  {Nelemans} G.,   {Portegies Zwart} S.,  2012, \mn@doi [\aap]
  {10.1051/0004-6361/201218966}, \href
  {http://adsabs.harvard.edu/abs/2012A%26A...546A..70T} {546, A70}

\bibitem[\protect\citeauthoryear{{Tutukov}, {Yungelson}  \&
  {Klayman}}{{Tutukov} et~al.}{1973}]{Tutukov1973}
{Tutukov} A.,  {Yungelson} L.,   {Klayman} A.,  1973, Nauchnye Informatsii,
  \href {http://adsabs.harvard.edu/abs/1973NInfo..27....3T} {27, 3}

\bibitem[\protect\citeauthoryear{{Ugliano}, {Janka}, {Marek}  \&
  {Arcones}}{{Ugliano} et~al.}{2012}]{Ugliano2012}
{Ugliano} M.,  {Janka} H.-T.,  {Marek} A.,   {Arcones} A.,  2012, \mn@doi
  [\apj] {10.1088/0004-637X/757/1/69}, \href
  {http://adsabs.harvard.edu/abs/2012ApJ...757...69U} {757, 69}

\bibitem[\protect\citeauthoryear{{Vink}}{{Vink}}{2016}]{Vink2016}
{Vink} J.~S.,  2016, preprint, \href
  {http://adsabs.harvard.edu/abs/2016arXiv161000578V} {} (\mn@eprint {arXiv}
  {1610.00578})

\bibitem[\protect\citeauthoryear{{Vink} \& {Gr{\"a}fener}}{{Vink} \&
  {Gr{\"a}fener}}{2012}]{Vink2012}
{Vink} J.~S.,  {Gr{\"a}fener} G.,  2012, \mn@doi [\apjl]
  {10.1088/2041-8205/751/2/L34}, \href
  {http://adsabs.harvard.edu/abs/2012ApJ...751L..34V} {751, L34}

\bibitem[\protect\citeauthoryear{{Vink} \& {de Koter}}{{Vink} \& {de
  Koter}}{2005}]{VinkdeKoter2005}
{Vink} J.~S.,  {de Koter} A.,  2005, \mn@doi [\aap]
  {10.1051/0004-6361:20052862}, \href
  {http://adsabs.harvard.edu/abs/2005A%26A...442..587V} {442, 587}

\bibitem[\protect\citeauthoryear{{Vink}, {de Koter}  \& {Lamers}}{{Vink}
  et~al.}{1999}]{Vink1999}
{Vink} J.~S.,  {de Koter} A.,   {Lamers} H.~J.~G.~L.~M.,  1999, \aap, \href
  {http://adsabs.harvard.edu/abs/1999A%26A...350..181V} {350, 181}

\bibitem[\protect\citeauthoryear{{Vink}, {de Koter}  \& {Lamers}}{{Vink}
  et~al.}{2001}]{Vink2001}
{Vink} J.~S.,  {de Koter} A.,   {Lamers} H.~J.~G.~L.~M.,  2001, \mn@doi [A \&
  {A}] {10.1051/0004-6361:20010127}, \href
  {http://adsabs.harvard.edu/abs/2001A%26A...369..574V} {369, 574}

\bibitem[\protect\citeauthoryear{{Vink}, {Muijres}, {Anthonisse}, {de Koter},
  {Gr{\"a}fener}  \& {Langer}}{{Vink} et~al.}{2011}]{Vink2011}
{Vink} J.~S.,  {Muijres} L.~E.,  {Anthonisse} B.,  {de Koter} A.,
  {Gr{\"a}fener} G.,   {Langer} N.,  2011, \mn@doi [\aap]
  {10.1051/0004-6361/201116614}, \href
  {http://adsabs.harvard.edu/abs/2011A%26A...531A.132V} {531, A132}

\bibitem[\protect\citeauthoryear{{Woosley}}{{Woosley}}{2017}]{Woosley2017}
{Woosley} S.~E.,  2017, preprint, \href {https://arxiv.org/abs/1608.08939} {}
  (\mn@eprint {arXiv} {1608.08939})

\bibitem[\protect\citeauthoryear{{Woosley}, {Blinnikov}  \& {Heger}}{{Woosley}
  et~al.}{2007}]{Woosley2007}
{Woosley} S.~E.,  {Blinnikov} S.,   {Heger} A.,  2007, \mn@doi [\nat]
  {10.1038/nature06333}, \href
  {http://adsabs.harvard.edu/abs/2007Natur.450..390W} {450, 390}

\bibitem[\protect\citeauthoryear{{Xu} \& {Li}}{{Xu} \& {Li}}{2010}]{Xi2010}
{Xu} X.-J.,  {Li} X.-D.,  2010, \mn@doi [\apj] {10.1088/0004-637X/716/1/114},
  \href {http://adsabs.harvard.edu/abs/2010ApJ...716..114X} {716, 114}

\bibitem[\protect\citeauthoryear{{Yoshida}, {Takahashi}, {Umeda}  \&
  {Ishidoshiro}}{{Yoshida} et~al.}{2016}]{Yoshida2016}
{Yoshida} T.,  {Takahashi} K.,  {Umeda} H.,   {Ishidoshiro} K.,  2016, \mn@doi
  [\prd] {10.1103/PhysRevD.93.123012}, \href
  {http://adsabs.harvard.edu/abs/2016PhRvD..93l3012Y} {93, 123012}

\bibitem[\protect\citeauthoryear{{Ziosi}, {Mapelli}, {Branchesi}  \&
  {Tormen}}{{Ziosi} et~al.}{2014}]{Ziosi2014}
{Ziosi} B.~M.,  {Mapelli} M.,  {Branchesi} M.,   {Tormen} G.,  2014, \mn@doi
  [\mnras] {10.1093/mnras/stu824}, \href
  {http://adsabs.harvard.edu/abs/2014MNRAS.441.3703Z} {441, 3703}

\bibitem[\protect\citeauthoryear{{de Mink} \& {Belczynski}}{{de Mink} \&
  {Belczynski}}{2015}]{DeMink2015}
{de Mink} S.~E.,  {Belczynski} K.,  2015, \mn@doi [\apj]
  {10.1088/0004-637X/814/1/58}, \href
  {http://adsabs.harvard.edu/abs/2015ApJ...814...58D} {814, 58}

\bibitem[\protect\citeauthoryear{{de Mink} \& {Mandel}}{{de Mink} \&
  {Mandel}}{2016}]{Demink2016}
{de Mink} S.~E.,  {Mandel} I.,  2016, \mn@doi [\mnras] {10.1093/mnras/stw1219},
  \href {http://adsabs.harvard.edu/abs/2016MNRAS.460.3545D} {460, 3545}

\makeatother
\end{thebibliography}




\appendix

\section{Core-collapse SNe}
\label{sec:appA}
In the following we summarize the main features of the rapid and delayed core-collapse SN mechanisms proposed by \citet{Fryer2012}.

\noindent In both cases, compact objects form from a proto-compact object $M_{\mathrm{pro}}$ that accretes mass from the fallback material $M_{\mathrm{fb}}$ which can follow the SN explosion,
\begin{equation}
	M_{\mathrm{fb}} =f_{\rm{fb}}(M_{\mathrm{fin}} - M_{\mathrm{pro}})~, 
\end{equation}		
where $M_{\mathrm{fin}}$ is the final mass of the star and $f_{\rm{fb}}$ is the fallback factor.
Starting from the baryonic mass of the compact object, $M_{\mathrm{rem,bar}} = M_{\mathrm{pro}} + M_{\mathrm{fb}}$, and considering the mass loss due to neutrinos it possible to compute the gravitational mass $M_{\mathrm{rem,grav}}$. We use the formula suggested by \citet{Timmes1996} for the NSs 
\begin{equation}
M_{\mathrm{rem,grav}} = \frac{\sqrt{1+0.3M_{\mathrm{rem,bar}}}-1}{0.15}~,	
\end{equation}
and the same approach described in \citet{Fryer2012} for BHs,
\begin{equation}
M_{\mathrm{rem,grav}} = 0.9 M_{\mathrm{rem,bar}}~. 	
\end{equation}

\subsubsection{Rapid}
For the rapid mechanisms, it is assumed a fixed mass of the proto-object, $M_{\mathrm{pro}} = 1.0$ \msun. The value of the fallback factor depends on the mass of the CO core $M_{\mathrm{CO}}$ and is given by
\begin{equation}
	f_{\mathrm{fb}} = \begin{cases} \frac{0.2}{M_{\mathrm{fin}} - M_{\mathrm{pro}}} & \rm{if}~~ M_{\mathrm{CO}}/\msun < 2.5  \cr
	\frac{0.286M_{\mathrm{CO}}-0.514\msun}{M_{\mathrm{fin}} - M_{\mathrm{pro}}} & \rm{if}~~ 2.5 \leq M_{\mathrm{CO}}/\msun < 6.0\cr
	1.0 & \rm{if}~~ 6.0 \leq M_{\mathrm{CO}}/\msun < 7.0\cr
	\alpha_{\rm R} M_{\mathrm{CO}} + \beta_{\rm R} & \rm{if}~~ 7.0 \leq M_{\mathrm{CO}}/\msun < 11.0\cr
	1.0 & \rm{if}~~11.0 \leq M_{\mathrm{CO}}/\msun,  \cr 
 	\end{cases}
\end{equation} 
where
\begin{equation}
	\alpha_{\rm R} \equiv 0.25 - \frac{1.275}{M_{\mathrm{fin}}- M_{\mathrm{pro}}} \\
	\beta_{\rm R} \equiv 1 - 11\alpha_{\rm R}~.
\end{equation}
The direct collapse of a star into a BH occurs when $f_{\mathrm{fb}}=1.0$ and for the rapid model it is verified in two intervals of core masses, $6.0\msun \leq M_{\mathrm{CO}} < 7.0 \msun$ and $11.0\msun \leq M_{\mathrm{CO}}$

\subsubsection{Delayed}
For the delayed model, even the mass of the proto-compact object depends on $M_{\mathrm{CO}}$ and it is given by,
\begin{equation}
	M_{\mathrm{pro}} = \begin{cases} 1.2\msun & \rm{if}~~ M_{\mathrm{core}}/\msun < 2.5  \cr
	1.3\msun & \rm{if}~~ 3.5 \leq M_{\mathrm{CO}}/\msun < 6.0\cr
	1.4\msun & \rm{if}~~ 6.0 \leq M_{\mathrm{CO}}/\msun < 11.0\cr
	1.6\msun & \rm{if}~~ 11.0 \leq M_{\mathrm{CO}}/\msun. \cr
 	\end{cases}
\end{equation} 
The fallback factor is computed by using the following expressions
\begin{equation}
	f_{\mathrm{fb}} = \begin{cases} \frac{0.2}{M_{\mathrm{fin}} - M_{\mathrm{pro}}} & \rm{if}~~ M_{\mathrm{CO}}/\msun < 2.5  \cr
	\frac{0.5M_{\mathrm{CO}}-1.05\msun}{M_{\mathrm{fin}} - M_{\mathrm{pro}}} & \rm{if}~~ 2.5 \leq M_{\mathrm{CO}}/\msun < 3.5\cr
	\alpha_{\rm D} M_{\mathrm{CO}} + \beta_{\rm D} & \rm{if}~~ 3.5 \leq M_{\mathrm{CO}}/\msun < 11.0\cr
	1.0 & \rm{if}~~11.0 \leq M_{\mathrm{CO}}/\msun,  \cr 
 	\end{cases}
\end{equation} 
where
\begin{equation}
	\alpha_{\rm D} \equiv 0.133 - \frac{0.093}{M_{\mathrm{fin}}- M_{\mathrm{pro}}} \\
	\beta_{\rm D} \equiv 1 - 11\alpha_{\rm D}~.
\end{equation}
Thus, for the delayed model the direct collapse of a star into a BH occurs only if $11.0\msun \leq M_{\mathrm{CO}}$.

\section{PISNe \& PPISNe}
\label{sec:appB}
\begin{figure}
	\centering
	\includegraphics[scale=0.3]{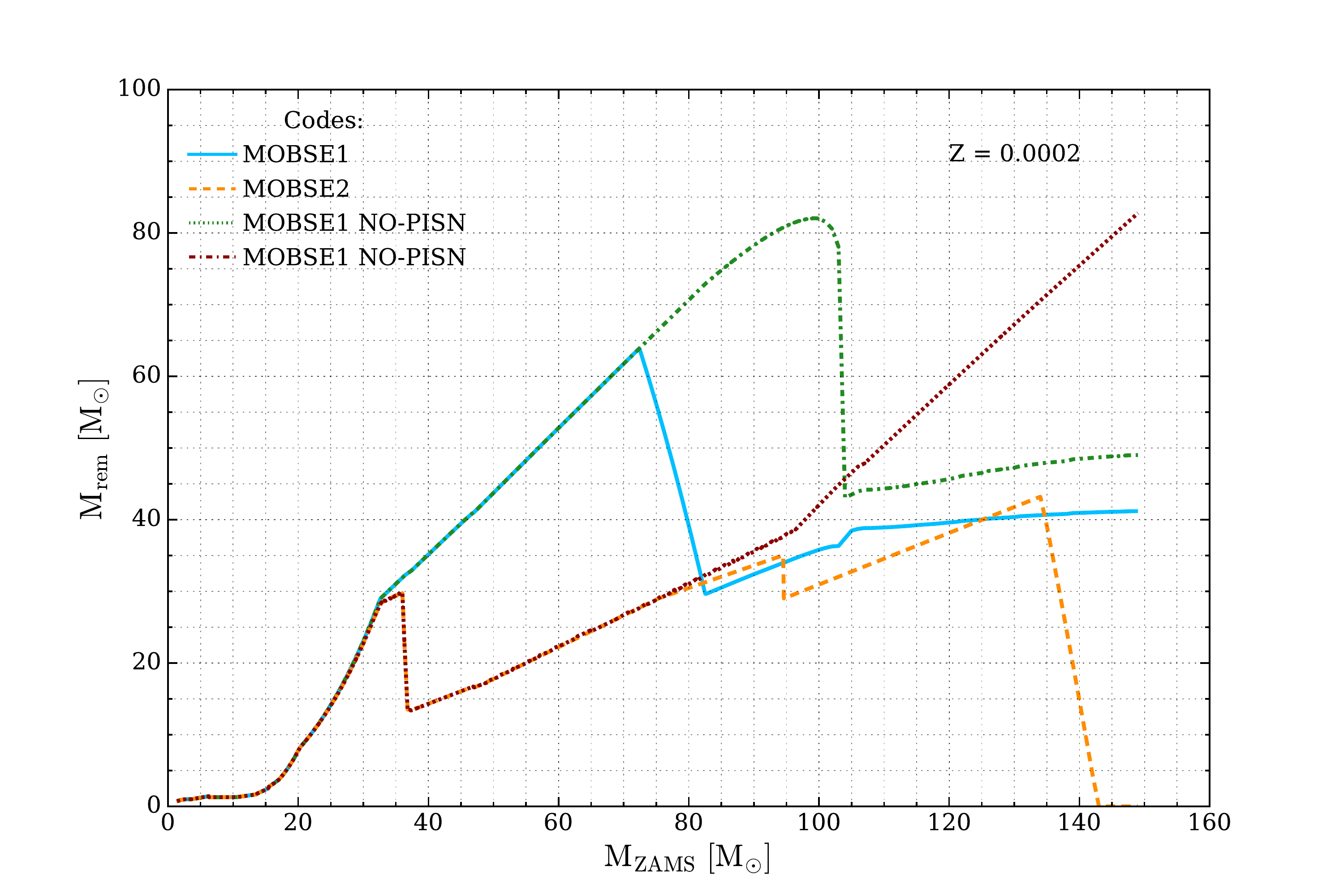}
	\caption{Effect of the PISNe and PPISNe on the mass spectrum of the remnants as a function of the $M_{\rm{ZAMS}}$ at $Z=0.0002$. Solid-blue line: \mobse1 with PISNe and PPISNe; dash-dot-green line: \mobse1 without PISNe and PPISNe; dashed-yellow line: \mobse2 with PISNe and PPISNe;  dotted-red line: \mobse2 without PISNe and PPISNe.    \label{fig:PPISN}}
\end{figure}
 In the following we detail the formulas implemented in \mobse{} to describe   PPISNe and   PISNe, following the the prescriptions described in \citet{Spera2016} and \citet{Spera2017}.  

To compute the mass of the compact remnant we adopted the formula 
\begin{equation}
	M_{\rm{rem}} = f_{\rm{p}} M_{\rm{rem,nop}},
\end{equation}
where $M_{\rm{rem,nop}}$ is the mass of the compact remnant we would obtain without PPISNe and PISNe and $f_{\rm{p}}$ is a factor depending on the final Helium core mass of the star. In particular, $f_{\rm{p}} = 1$ means that remnants will form via direct collapse and $f_{\rm{p}} = 0$ means that remnants will completely destroy due to PISNe. In Fig. \ref{fig:PPISN} we show the mass spectrum with/without PPISNe and PISNe for both \mobse1 and \mobse2 at $Z=0.0002$.

We use the following expressions to compute $f_{\rm{p}}$ distinguishing between H-rich stars and WR stars. 
\subsubsection{Normal stars}
\begin{equation}
	f_{\mathrm{p}} = 
	\begin{cases} 
		1.0 & \rm{if}~~ M_{\mathrm{He}}/\msun \leq 32  \cr
		\frac{(k - 1.0)}{5.0}M_{\mathrm{He}} + \frac{(37.0 - 32.0\alpha_{\rm{n}})}{5.0} 	  & \rm{if}~~ 32 < M_{\mathrm{He}}/\msun \leq 37 \cr
		\alpha_{\rm{n}} & \rm{if}~~ 37 < M_{\mathrm{He}}/\msun \leq 60 \cr
		-\frac{\alpha_{\rm{n}}}{4.0}	M_{\mathrm{He}} +16.0\alpha_{\rm{n}} & \rm{if}~~ 60 < M_{\mathrm{He}}/\msun < 64 \cr
		0.0 & \rm{if}~~ 64 \leq M_{\mathrm{He}}/\msun < 135 \cr
		1.0 & \rm{if}~~ 135 \leq M_{\mathrm{He}}/\msun, \cr
 	\end{cases}
\end{equation}
where $\alpha_{\rm{n}}$ is given by
\begin{equation}
	\alpha_{\rm{n}} = 0.67 \frac{M_{\mathrm{He}}}{M_{\mathrm{tot}}}  + 0.1.
\end{equation} 
\subsubsection{WR stars}
\begin{equation}
	f_{\mathrm{p}} = 
	\begin{cases} 
		1.0 \cr 
		\qquad \rm{if}~~ M_{\mathrm{He}}/\msun \leq 32  \cr
		(M_{\mathrm{He}} - 32.0)(0.5226\frac{M_{\mathrm{He}}}{M_{\mathrm{tot}}} -0.52974) + 1.0 	 \cr 
		\qquad  \rm{if}~~ 32 < M_{\mathrm{He}}/\msun \leq 37 \cr
		(- 0.1381\frac{M_{\mathrm{He}}}{M_{\mathrm{tot}}} + 0.1309)(M_{\mathrm{He}} - 56) + 0.82916 \cr 
		\qquad  \rm{if}~~ 37 < M_{\mathrm{He}}/\msun \leq 56 \cr
		- 0.103645M_{\mathrm{He}} + 6.63328 \cr 
		\qquad  \rm{if}~~ 56 < M_{\mathrm{He}}/\msun < 64 \cr
		0.0			  \cr 
		\qquad \rm{if}~~ 64 \leq M_{\mathrm{He}}/\msun < 135 \cr
		1.0			  \cr 
		\qquad  \rm{if}~~ 135 \leq M_{\mathrm{He}}/\msun. \cr
 	\end{cases}
\end{equation}

\bsp	
\label{lastpage}
\end{document}